\documentclass{article} 
\usepackage{styles,times}
\usepackage{lipsum}
\usepackage{multirow}
\usepackage{booktabs}
\usepackage{multirow}
\usepackage{multicol}
\usepackage{adjustbox}
\usepackage{tcolorbox}
\usepackage{tabularx}
\usepackage{tikz}
\usepackage{graphicx}

\usepackage{amsmath,amsfonts,bm}

\def\eqref#1{equation~\ref{#1}}

\def\1{\bm{1}}

\DeclareMathAlphabet{\mathsfit}{\encodingdefault}{\sfdefault}{m}{sl}
\SetMathAlphabet{\mathsfit}{bold}{\encodingdefault}{\sfdefault}{bx}{n}

\usepackage{hyperref}
\usepackage{url}
\usepackage{lineno}
\usepackage{array, makecell}
\usepackage{algorithm}
\usepackage{algpseudocode}
\usepackage[numbers]{natbib}
\usepackage{geometry}
\usepackage{amssymb}
\usepackage{chngcntr}
\usepackage{subcaption}
\usepackage{enumitem}
\usepackage{overpic}

\usepackage{float} 
\title{\textsc{Electronic Structure Guided Inverse Design Using Generative Models}}

\author{Shuyi Jia$^{\dag}$, Panchapakesan Ganesh$^{\ddag\text{\textsection}}$, Victor Fung$^{\dag\text{\textsection}}$ \\
{\normalsize $^\dag$Computational Science and Engineering, Georgia Institute of Technology, Atlanta, GA, USA} \\
{\normalsize $^\ddag$Center For Nanophase Materials Sciences, Oak Ridge National Laboratory, Oak Ridge, TN, USA}\\
{\normalsize $^{\text{\textsection}}$Corresponding authors: 
\texttt{ganeshp@ornl.gov}, 
\texttt{victorfung@gatech.edu}}
}
\date{}

\newcommand{\bv}[1]{\boldsymbol{#1}}

\newcolumntype{Y}{>{\centering\arraybackslash}X}

\iclrfinalcopy 
\begin{document}

\maketitle

\begin{abstract}
The electronic structure of a material fundamentally determines its underlying physical, and by extension, its functional properties. Consequently, the ability to identify or generate materials with desired electronic properties would enable the design of tailored functional materials. Traditional approaches relying on human intuition or exhaustive computational screening of known materials remain inefficient and resource-prohibitive for this task. Here, we introduce DOSMatGen, the first instance of a machine learning method which generates crystal structures that match a given desired electronic density of states. DOSMatGen is an E(3)-equivariant joint diffusion framework, and utilizes classifier-free guidance to accurately condition the generated materials on the density of states. Our experiments find this approach can successfully yield materials which are both stable and match closely with the desired density of states. Furthermore, this method is highly flexible and allows for finely controlled generation which can target specific templates or even individual sites within a material. This method enables a more physics-driven approach to designing new materials for applications including catalysts, photovoltaics, and superconductors.

\end{abstract}

\section{Introduction}\label{sec:intro}
The ability to design novel materials with specific functional properties is fundamental to technological breakthroughs, driving innovation in fields such as energy storage, electronics, and healthcare \citep{deringer2020modelling, liu2020two, lou2020reviews, iqbal2021advances, ling2022review}. While traditional approaches to materials design rely heavily on Edisonian trial-and-error testing, or exhaustive screening through experimental or computational methods, these methods are often both time-consuming and resource-intensive \citep{pyzer2015high, liu2017materials}. In contrast, recent advances in deep generative models have enabled them to be a potentially transformative approach to materials design. Specifically, this approach to inverse design eliminates the need for exhaustive screening of the materials space by working backwards to create materials from the ground up via a data-driven process \citep{bilodeau2022generative, noh2020machine, fuhr2022deep, park2024has}.


Deep generative models for materials design can be trained to conditionally generate the precise crystal structure of materials, which allows for further computational investigation followed by eventual experimental synthesis and validation. Current state-of-the-art approaches generally operate by generating new candidates within a given materials representation, or by applying an iterative process such as denoising directly on atomic coordinates in the Cartesian space. Examples of the former case include text-based representations for autoregressive language modeling of materials \citep{flam2023language, antunes2023crystal, gruver2024fine}, 3D voxelized images for variational autoencoders and generative adversarial networks (GANs) \citep{hoffmann2019data, court20203, long2021constrained}, among others \citep{fung2022atomic, yang2023scalable, sinha2024representation, joshi2025all}. In the latter case, a seminal early example would be the crystal diffusion variational autoencoder (CDVAE) \citep{xie2021crystal}, which employs a diffusion model \citep{ho2020denoising} to denoise atomic coordinates while predicting composition and lattice parameters from a latent representation. Building on this foundation, recent diffusion-based and flow-matching approaches have advanced to jointly generate all relevant structural and compositional attributes in a unified process \citep{jiao2024crystal, zeni2025generative, miller2024flowmm}. Using these methods, new materials have been generated with desired properties which have been validated with density functional theory (DFT) \citep{zhao2023physics, ren2022invertible, wines2023inverse, kwon2024spectroscopy} and in some cases, experimental studies \citep{zeni2025generative}.

\begin{figure}[h]
\centering
\begin{tikzpicture}
    \draw (0, 0) node[inner sep=0] {\includegraphics[width=\textwidth]{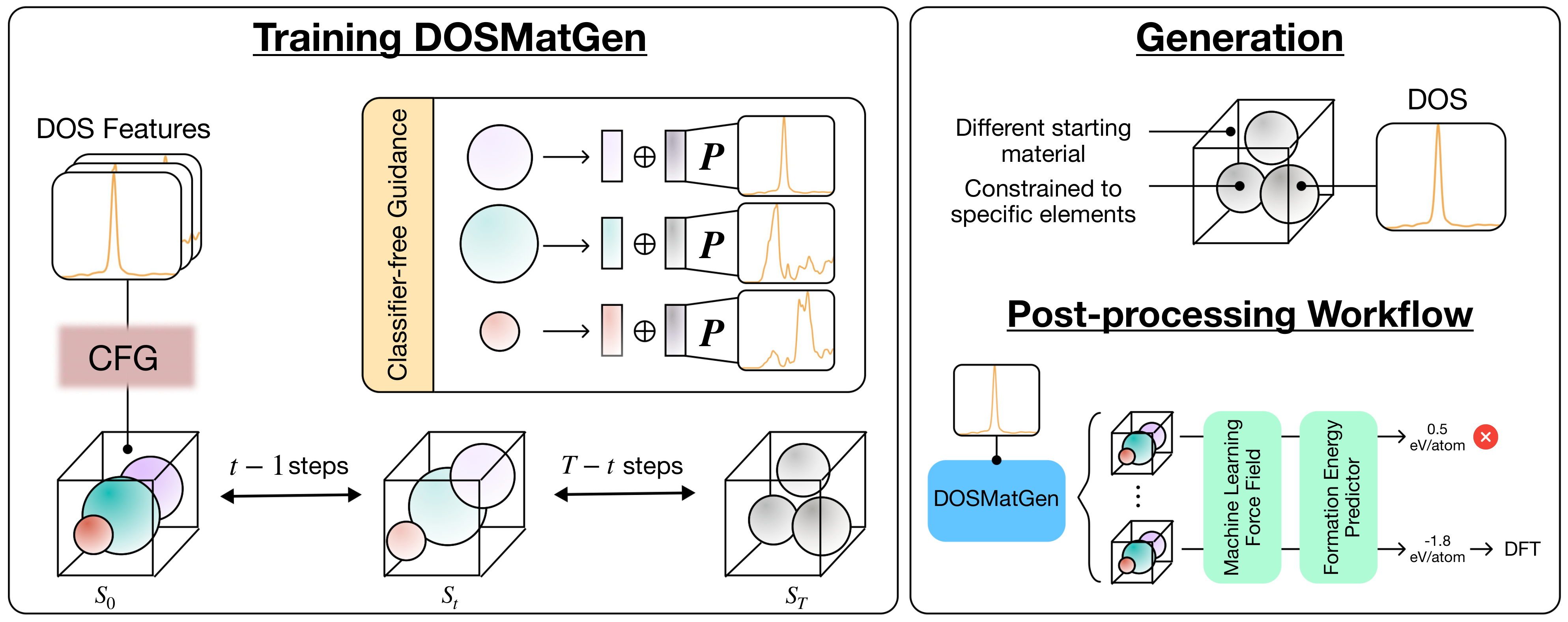}};
\end{tikzpicture}
\caption{Overview of DOSMatGen. The training process of the denoising neural network in DOSMatGen employs classifier-free guidance (CFG). A projection layer $\boldsymbol{P}$ maps node-level density of states (DOS) features into the same space as the node embeddings, which are then combined through element-wise addition. During training, a structure $S_0$ is progressively \textit{noisified} to $S_t$ at timestep $t$ and passed into the network to predict the noise added. During generation, DOSMatGen can condition one or more atoms on target DOS features, ensuring the denoised structure aligns with the specified conditions. Additional constraints, such as fixing element types or using structural templates as starting points, can be incorporated for finer control over the generation process. Furthermore, DOSMatGen can be integrated as part of larger materials discovery workflows, generating large sets of candidate structures that are subsequently relaxed using a machine learning force field (MLFF) and evaluated with a formation energy predictor. Candidates below a specified formation energy threshold can then be filtered for further analysis, including DFT validation.}
\label{fig:overview}
\end{figure}

In this work, we introduce DOSMatGen, a conditional diffusion model designed to generate stable crystalline materials with targeted atomic-level electronic properties. Extending the periodic E(3)-equivariant denoising framework of DiffCSP \citep{jiao2024crystal}, DOSMatGen achieves high-quality conditional generation by carefully tuning the balance between the structural validity of the generated materials and their alignment with the target properties. As a concrete demonstration, we focus on conditioning the diffusion process on node-level density of states (DOS). The DOS corresponds to a 1D spectral property that encodes the key electronic-level functionality of a material, from its metal, semiconducting, insulating states, to more exotic quantum phases such as superconductivity and topological quantum states. Prior deep-learning studies have yielded models which can predict the DOS from a given atomic structure \citep{kong2022density, fung2022physically}, but the \textit{inverse process} of generating materials matching a given DOS remains highly challenging.
To effectively utilize the DOS information as a condition for structure generation, we implemented and evaluated two distinct guidance strategies for diffusion models: classifier-based guidance \citep{dhariwal2021diffusion} and classifier-free guidance \citep{ho2022classifier}. We find classifier-free guidance outperforms classifier-based guidance, offering a more robust mechanism for aligning generated structures with the desired properties. Next, using the classifier-free guidance, we demonstrate DOSMatGen's capability for fine-grained control over atomic structure generation through three constrained generation tasks that selectively impose distinct constraints on specific subsets of atoms within a single structure. Finally, we demonstrate a post-processing workflow to yield stable, optimized crystal structures which are ready for DFT validation. An overview of the DOSMatGen framework is shown in Fig. \ref{fig:overview}.

\section{Methodology}\label{sec:method}
\subsection{Representation of Periodic Materials}
A periodic material is usually represented by its unit cell---the smallest repeating arrangement of atoms in Cartesian coordinates that, when translated in space, generates the complete crystal structure. Given a structure $S$ with $N$ atoms in its unit cell, it can be fully described by the tuple $S=\left( \mathbf{A}, \mathbf{X}, \mathbf{L} \right)$, where $\mathbf{L} = \left(\bv{l}_1,\bv{l}_2, \bv{l}_3\right) \in \mathbb{R}^{3\times 3}$ is a matrix of lattice vectors, $\mathbf{X} = \left( \bv{x}_1,\ldots, \bv{x}_N\right) \in \mathbb{R}^{N\times 3}$ represents the Cartesian coordinates of the $N$ atoms within the unit cell, and $\mathbf{A}\in \left\{ 0,1 \right\}^{N\times Z}$ is a matrix of one-hot encoded representations of $Z$ possible elements for each atom. Note that the unit cell can be translated arbitrary in space by $\mathbf{L}$. In this work, we use fractional coordinates, obtained by transforming Cartesian coordinates $\bv{x}$ through the inverse lattice matrix: $\bv{x}_f = \mathbf{L}^{-1} \bv{x}$. For notational clarity, we omit the subscript $_f$ when the coordinate system is contextually clear.

\subsection{Joint Diffusion Model}
Given a structure $S = \left( \mathbf{A}, \mathbf{X}, \mathbf{L}\right)$, DOSMatGen simultaneously diffuses the atom types $\mathbf{A}$, the fractional coordinate matrix $\mathbf{X}$ and the lattice $\mathbf{L}$. Given a timestep $t \in \left[0, T\right]$, $S_t = \left( \mathbf{A}_t, \mathbf{X}_t, \mathbf{L}_t\right)$ denotes the intermediate \textit{noisy} structure. In particular, the forward diffusion process deterministically adds noise to $S_0$, while the backward denoising process iteratively samples from the prior distribution $S_T$ to recover the original structure $S_0$. If the prior distribution $p(S_T)$ is invariant and the Markov transition $p(S_{t-1}|S_t)$ is equivariant, then the recovered distribution from $S_T$ will be periodic E(3)-invariance \citep{xu2022geodiff}.  The overall objective for training the joint diffusion model on $\mathbf{A}$, $\mathbf{X}$ and $\mathbf{L}$ is
\begin{align}
    \mathcal{L}_S = \lambda_{\mathbf{A}}\mathcal{L}_{\mathbf{A}}+ \lambda_{\mathbf{X}}\mathcal{L}_{\mathbf{X}} + \lambda_{\mathbf{L}}\mathcal{L}_{\mathbf{L}}, \label{eq:combined-training-obj}
\end{align}
where $\lambda$ represents the corresponding weight given to each individual loss component. The full diffusion objectives can be found in Appendix \ref{sec:appendix-a}.

\subsection{Classifier-based Guidance}

Guided diffusion models typically generate structures $S$ from the conditional distribution $p(S|y)$, where $y$ represents the target label or condition. In classifier-based guidance \cite{dhariwal2021diffusion}, this is achieved by leveraging Bayes' Rule to guide the diffusion process:
\begin{align}
\nabla_{S_t} \log p_t (S_t|y) = \nabla_{S_t} \log p_t(S_t) + \nabla_{S_t} \log p_t(y|S_t), \label{eq:classifer-guidance}
\end{align}

where the gradient of the log conditional probability $\nabla_{S_t} \log p_t (S_t|y)$ at timestep $t$ given the condition $y$ can be decomposed into two terms: $\nabla_{S_t} \log p_t(S_t)$ and $\nabla_{S_t} \log p_t(y|S_t)$. The first term corresponds to the unconditional generative model, which guides the generation of $S_t$ and is generally approximated via a neural network $f_\theta (S_t,t)$. The second term represents the gradient from a time-dependent classifier, which estimates how likely $S_t$ is to match the condition $y$ and is approximated by a separate neural network $g_\phi (y|S_t, t)$. In practice, because the unconditional model and the classifier are separate neural networks, we train them independently. 

\subsection{Classifier-free Guidance}
In contrast to classifier-based guidance, classifier-free guidance \cite{ho2022classifier} integrates the conditional generation directly into the diffusion model by training a single neural network for both conditional and unconditional models. Instead of learning the unconditional score function $\nabla_{S_t}\log p_t (S_t)$ independently, classifier-free guidance learns the unconditional and conditional distributions jointly:
\begin{align}
     \nabla_{S_t}\log p_t(S_t,c), \quad c = \begin{cases}
\varnothing & \text{ with probability } p_{\text{uncond}} \\
y & \text{ else }
\end{cases}
\end{align}
where $p_{\text{uncond}}$ is some probability for unconditional generation, set as a hyper-parameter. During sampling, the mixed score function becomes $(1+\omega)\nabla_{x_t}\log p_t (x_t, y) - \omega \nabla_{x_t}\log p_t (x_t, \varnothing)$, where $\omega$ is the guidance strength. In practice, we project $y$ into the same space as the node representations, and combine them through element-wise addition. In our experiments, we let $p_{\text{uncond}} = 0.2$ and $\omega = 1$.

\subsection{Dataset}
Our training dataset for the denoising neural network, which we henceforth refer to as ``MP DOS," comprises of 50,790 inorganic crystal structures curated from the Materials Project repository \citep{jain2013commentary}. The target property is the atom-level density of states (DOS), with each atom in every structure labeled by its corresponding DOS. The MP DOS dataset is divided into training, testing, and validation splits in a \texttt{0.6:0.2:0.2} ratio, with only the training split used to train the denoising network, as well as the forward model in classifier-based guidance.

\section{Results}\label{sec:results}

\subsection{Conditional Generation}
We first evaluate the two diffusion model guidance strategies—classifier-based and classifier-free using the testing split of the MP DOS dataset. At the initial timestep $t=T$ of generation, we specify the number of atoms and the DOS while randomly sampling atom types $\mathbf{A}_T$, fractional coordinates $\mathbf{X}_T$, and lattice matrix $\mathbf{L}_T$ from their respective prior distributions $p(\mathbf{A}_T)$, $p(\mathbf{X}_T)$ and $p(\mathbf{L}_T)$. In other words, our goal is to reconstruct the entire testing split of the MP DOS dataset using the atom-level DOS as the conditional input. In addition to $T=1000$, which represents generation from fully randomized initialized structures, we perform additional experiments by generating structures from timesteps $T=200$ and $T=500$. These cases involve partially \textit{noisifying} the original testing split structures using their forward noising processes (see Appendix \ref{sec:appendix-a}). As the timestep decreases, more structural information from the original data is retained, leading to generated structures that more closely resemble those in the testing split.

To compare the DOS of generated structures with the ground truth conditioning data, we trained a separate surrogate model that predicts atom-level DOS from input crystal structures. We first evaluate the conditional generation by calculating the mean absolute errors (MAEs) between generated structures' DOS and the ground truth DOS. In particular, we let $y$ denote the ground truth DOS used to condition the generation process. We define $\hat{y}$ as the predicted DOS from the ground truth structures, and 
$\hat{y}_{\text{gen}}$ as the predicted DOS from the generated structures. Table \ref{tab:conditional-gen-mae} presents the MAEs comparing: (1) $y$ with $\hat{y}_{\text{gen}}$, and (2) $\hat{y}$ with $\hat{y}_{\text{gen}}$. We observe that both $\text{MAE}\left(\hat{y}_{\text{gen}}, y\right)$ and $\text{MAE}\left(\hat{y}_{\text{gen}}, \hat{y}\right)$ increase as the starting timestep $T$ of the diffusion process becomes greater. This aligns with the fact that as $T$ increases, less information of the original crystal structures is retained, making the diffusion process more reliant on generating high-quality structures from pure noise. Consequently, the generated structures are more prone to deviate further from the ground truth, leading to higher errors in the predicted DOS. 

When comparing classifier-based and classifier-free guidance, we observe from Table \ref{tab:conditional-gen-mae} that classifier-free guidance consistently achieves better performance. The difference in error between the two strategies is relatively small at $T=200$, which is expected, as the model retains enough information from the original ground truth structures to enable both models to denoise more effectively to the ground truth structures, and resulting in lower values of both $\text{MAE}\left(\hat{y}_{\text{gen}}, y\right)$ and $\text{MAE}\left(\hat{y}_{\text{gen}}, \hat{y}\right)$. However, at the starting timestep $T=1000$, where no information from the original structures is present, the performance of classifier-based guidance deteriorates significantly. Specifically, when compared to that at $T=200$, $\text{MAE}\left(\hat{y}_{\text{gen}}, y\right)$ at $T=1000$ increases from 0.114 to 0.268, representing a 135\% increase, while $\text{MAE}\left(\hat{y}_{\text{gen}}, \hat{y}\right)$ rises from 0.068 to 0.256, marking a 276\% increase. In contrast, classifier-free guidance exhibits significantly lower MAE increases from timestep $T=200$ to $T=1000$: a 17.6\% rise in $\text{MAE}\left(\hat{y}_{\text{gen}}, y\right)$ and a 79.6\% increase in $\text{MAE}\left(\hat{y}_{\text{gen}}, \hat{y}\right)$. In particular, the forward surrogate model's own error, 
$\text{MAE}\left(\hat{y}, y\right)$, is 0.096. This indicates that even at $T=1000$, classifier-free guidance generates structures that align with the target DOS reasonably well with an MAE of 0.120. Additionally, at $T=1000$, 314 out of 10,158 structures generated with classifier-based guidance fail to even map to a valid crystal graph for the forward surrogate model, highlighting its inferior performance compared to classifier-free guidance.



\begin{table}[H]
\centering
\resizebox{\textwidth}{!}{%
\setlength{\tabcolsep}{3.5pt} 
\renewcommand{\arraystretch}{1.3}
\begin{tabular}{@{}lccccccccc@{}}
\toprule
& \multicolumn{2}{c}{$T=200$} & \phantom{abc} & \multicolumn{2}{c}{$T=500$} & \phantom{abc} & \multicolumn{2}{c}{$T=1000$} \\
\cmidrule{2-3} \cmidrule{5-6} \cmidrule{8-9}
& $\text{MAE}\left(\hat{y}_{\text{gen}}, y\right)$ & $\text{MAE}\left(\hat{y}_{\text{gen}}, \hat{y}\right)$ && $\text{MAE}\left(\hat{y}_{\text{gen}}, y\right)$ & $\text{MAE}\left(\hat{y}_{\text{gen}}, \hat{y}\right)$ && $\text{MAE}\left(\hat{y}_{\text{gen}}, y\right)$ & $\text{MAE}\left(\hat{y}_{\text{gen}}, \hat{y}\right)$ \\
\midrule
Classifier      & 0.114 & 0.068 && 0.200 & 0.182 && $0.268^{\dagger}$           & $0.256^{\dagger}$ \\
Classifier-free & 0.102 & 0.049 && 0.109 & 0.065 && 0.120\phantom{$^{\dagger}$} & 0.088\phantom{$^{\dagger}$} \\
\bottomrule
\end{tabular}
}
\caption{
Mean absolute errors (MAEs) of density of states (DOS) comparing structures generated using classifier-based and classifier-free guidance. $\hat{y}_{\text{gen}}$ is the DOS predicted from generated structures, $\hat{y}$ is the DOS predicted from ground truth structures and $y$ is the target DOS from test set. Comparisons are made across three different denoising timesteps, $T \in \{200, 500, 1000\}$, with $T=1000$ representing generation from fully random initializations. $\dagger$: 314 out of 10,158 generated structures were excluded due to errors in constructing the crystal graph. The forward surrogate model's intrinsic error, $\text{MAE}\left(\hat{y}, y\right)$, is 0.096.
}
\label{tab:conditional-gen-mae}
\end{table}

\begin{table}[H]
\centering
\resizebox{\textwidth}{!}{%
\setlength{\tabcolsep}{3.2pt} 
\renewcommand{\arraystretch}{1.3}
\begin{tabular}{@{}lccccccccc@{}}
\toprule
& \multicolumn{2}{c}{$T=200$} & \phantom{abc} & \multicolumn{2}{c}{$T=500$} & \phantom{abc} & \multicolumn{2}{c}{$T=1000$} \\
\cmidrule{2-3} \cmidrule{5-6} \cmidrule{8-9}
& \footnotesize{Struct. Match \%} & \footnotesize{Comp. Match \%} && \footnotesize{Struct. Match \%} & \footnotesize{Comp. Match \%} && \footnotesize{Struct. Match \%} & \footnotesize{Comp. Match \%}  \\
\midrule
Classifier      & 63.5 & 73.8 && $4.23^{\dagger}$  & 6.01 && $0.07^{\ddagger}$  & 2.95 \\
Classifier-free & 81.6 & 94.4 && 42.0\phantom{$^{\dagger}$} & 73.0 && 14.7\phantom{$^{\ddagger}$} & 58.7 \\
\bottomrule
\end{tabular}
}
\caption{
Structure match rate and composition match rate for structures generated using classifier-based and classifier-free guidance, compared to the test set. The structure match rate is the percentage of structures satisfying the tolerances (\texttt{stol=0.5}, \texttt{angletol=10}, \texttt{ltol=0.3}) using pymatgen's \texttt{StructureMatcher}. The composition match rate is the percentage of structures whose compositions perfectly match those of the test set. Comparisons are made across three different denoising timesteps, $T \in \{200, 500, 1000\}$, with $T=1000$ representing generation from fully random initializations. $\dagger$/$\ddagger$: 9/372 out of 10,158 generated structures were excluded due to timeouts during metric calculation.
}
\label{tab:conditional-gen-match}
\end{table}

In Table \ref{tab:conditional-gen-match}, we evaluate the performance of the conditional generation using (1) structure match rate and (2) composition match rate. The structure match rate quantifies the percentage of generated structures that are geometrically similar to the ground truth structures, as determined by the \texttt{StructureMatcher} algorithm from \texttt{pymatgen} within specified tolerances \citep{xie2021crystal, ong2013python}. The composition match rate represents the percentage of generated structures that exactly match the ground truth structures in terms of chemical composition. From Table \ref{tab:conditional-gen-match}, we observe that as $T$ increases, both structure and composition match rates decreases, highlighting that higher amount of noise during the generation process result in structures that deviate further from the ground truth. Comparing the two guidance strategies, classifier-free guidance consistently demonstrates superior performance. At timestep $T=1000$, it maintains a structure match rate of 14.7\% and a composition match rate of 58.7\%, significantly outperforming classifier-based guidance's match rates of 0.07\% and 2.95\%, respectively.

\begin{figure}[h]
     \centering
     \begin{subfigure}[b]{0.485\textwidth}
         \centering
         \includegraphics[width=\textwidth]{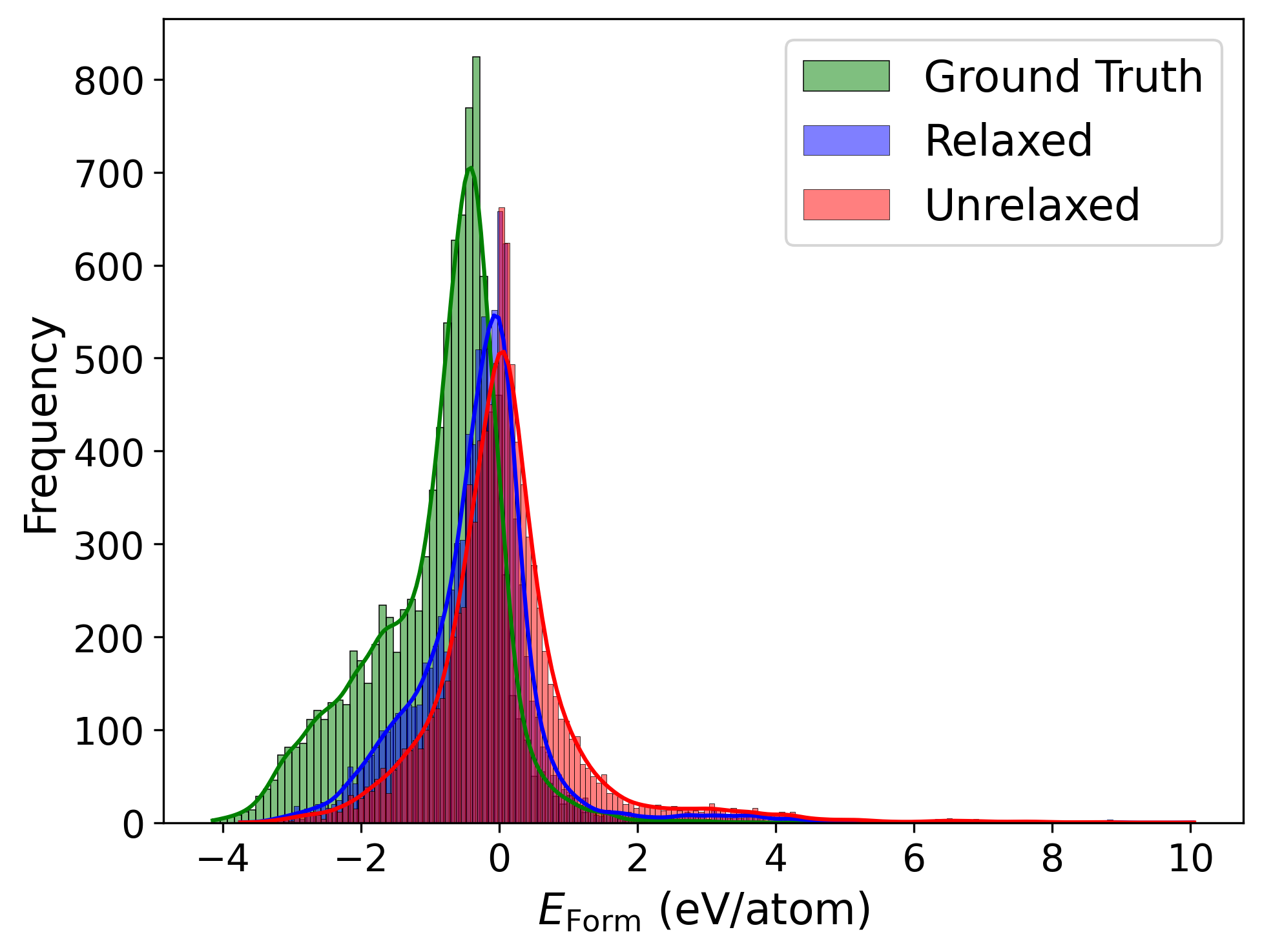}
         \caption{Classifier-based guidance}
         \label{fig:cg_distribution}
     \end{subfigure}
     \begin{subfigure}[b]{0.485\textwidth}
         \centering
         \includegraphics[width=\textwidth]{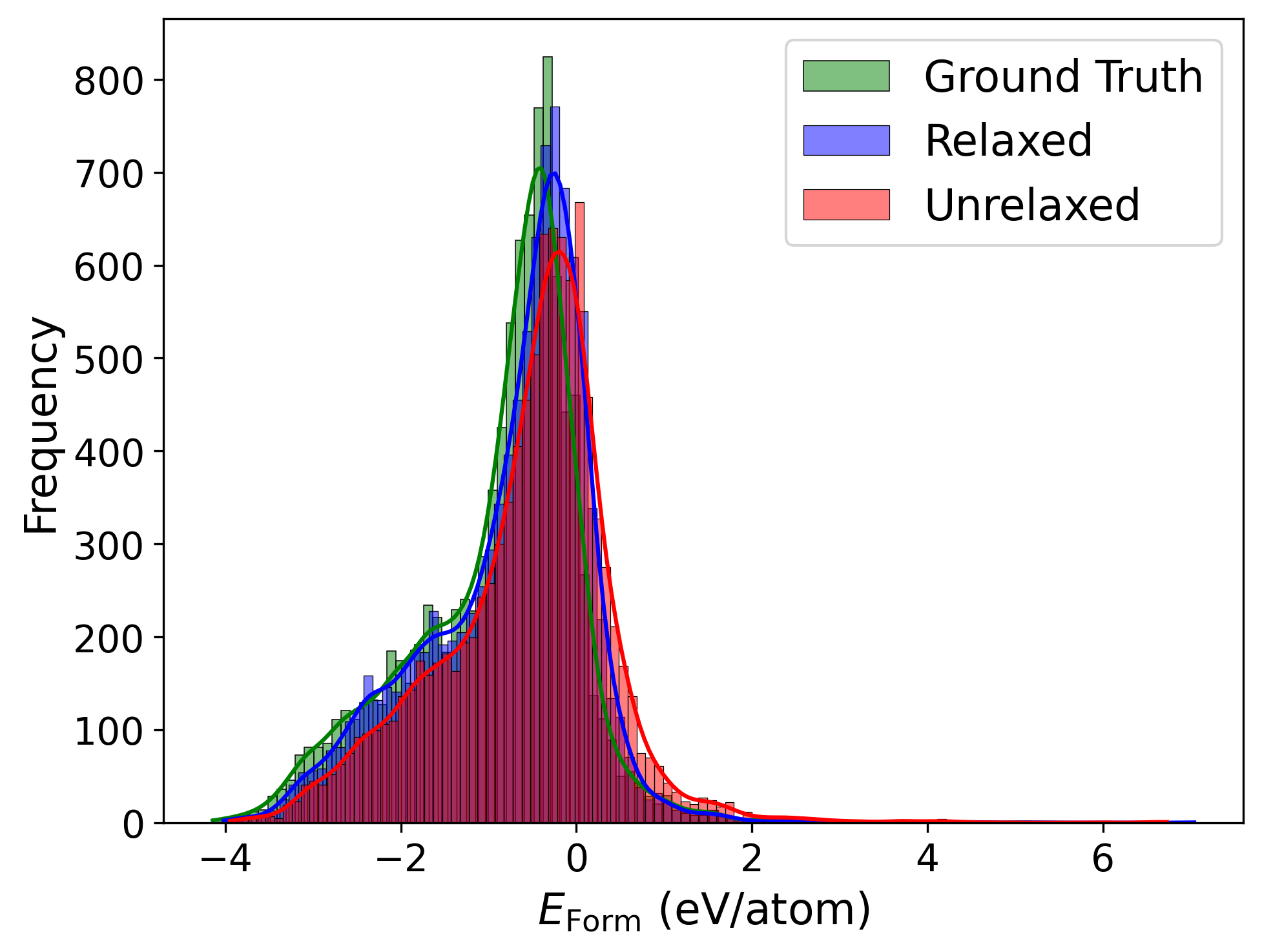}
         \caption{Classifier-free guidance}
         \label{fig:cfg_distribution}
     \end{subfigure}
        \caption{Distribution of formation energies per atom (eV/atom) for structures conditionally generated at $T=1000$ using (a) classifier-based guidance and (b) classifier-free guidance. All generated structures are relaxed using the \texttt{Orb-v2} interatomic potential. The ground truth distribution corresponds to the formation energies of the original test set structures, with all energies predicted using \texttt{Orb-v2}.}
        \label{fig:testset_distributions}
\end{figure}

In Fig. \ref{fig:testset_distributions}, we present the distribution of formation energies per atom for the structures generated in Table \ref{tab:conditional-gen-mae} at $T=1000$. Structure relaxation and formation energy calculations for all structures are performed using \texttt{Orb-v2} \citep{neumann2024orb}, an accurate and efficient universal machine learning force field (MLFF). The ground truth distribution corresponds to the original 10,158 structures from the test set. The results in Fig. \ref{fig:testset_distributions} show that structures generated with classifier-free guidance closely match the ground truth distribution of formation energies, exhibiting significant overlap. Notably, the majority of these structures---relaxed or unrelaxed---have formation energies below 0 eV/atom, indicating a reasonable range of thermodynamic stability. In contrast, structures generated with classifier-based guidance are noticeably shifted toward higher formation energies and deviate further from the ground truth distribution. The distribution plots for structures generated at $T=200$ and $T=500$ are provided in Appendix \ref{sec:appendix-b}.

From Table \ref{tab:conditional-gen-mae} and \ref{tab:conditional-gen-match}, it is clear that classifier-free guidance consistently outperforms classifier-based guidance. It achieves lower MAEs across all starting timesteps, and maintains significantly higher structure and composition match rates at $T=1000$. While classifier-free guidance excels in generating structures that align closely with the ground truth both structurally and compositionally, it is important to recognize that such strong alignment can be a double-edged sword. On one hand, it naturally leads to structures which match better to the desired DOS, which is central to our conditional generation approach. On the other hand, this convergence towards the ground truth structures may come at the cost of reduced diversity in the generated structures, which would be detrimental to the goal of finding novel materials. Based on our evaluation, we find the classifier-free version of DOSMatGen yields good agreement with the desired DOS while having a moderate match rate which is neither too high nor too low, making it suitable for materials discovery tasks. Next, we will apply DOSMatGen to three different types of constrained generation tasks to further assess the model. 



\subsection{Constrained Generation}
\begin{figure}[h]
\centering
\begin{tikzpicture}
    \draw (0, 0) node[inner sep=0] {\includegraphics[width=\textwidth]{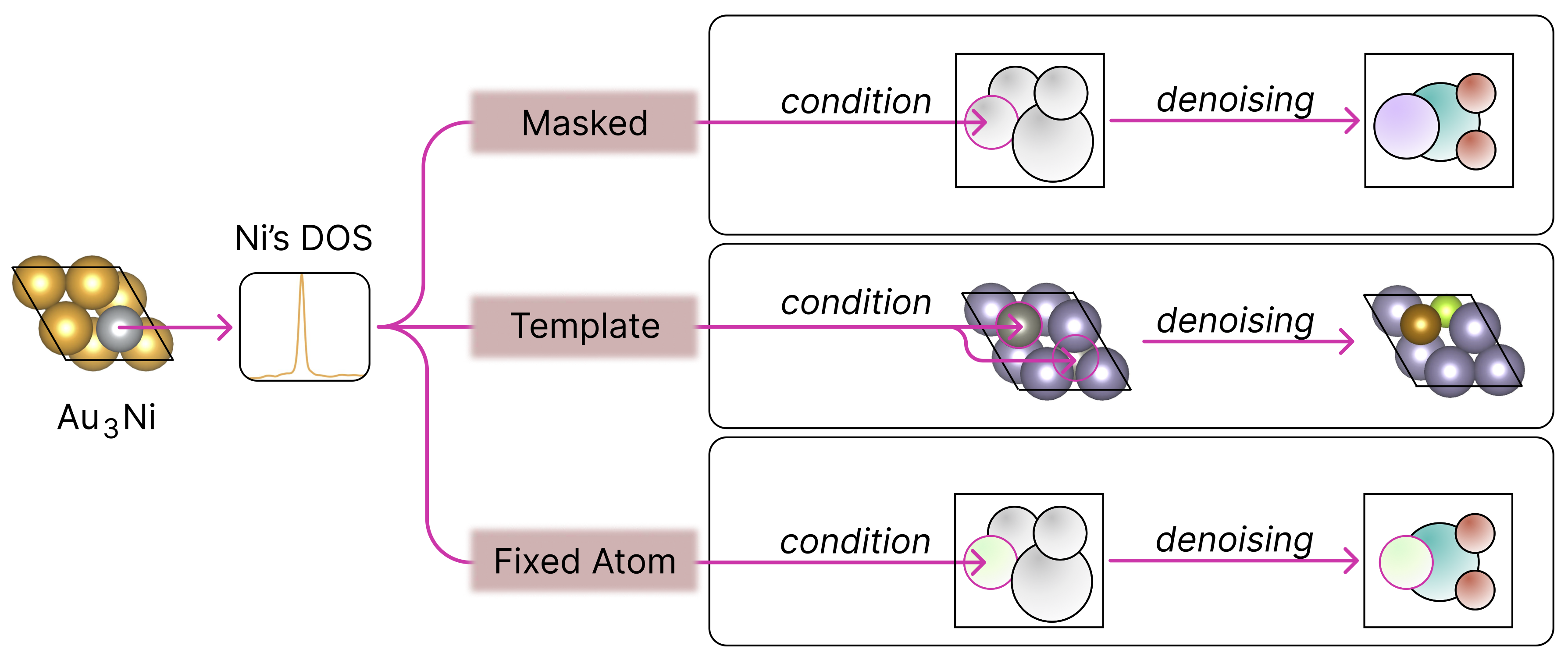}};
\end{tikzpicture}
\caption{A visual representation of the three constrained generation tasks. (1) Masked generation: a single atom of the structure is conditioned on Ni's DOS, while the remaining atoms are left unconditioned. (2) Template guided generation: Instead of complete random initialization, use an existing material as a template. A subset of the template material's atoms is then conditioned on Ni's DOS. (3) Fixed atom type generation: a single atom of the structure is conditioned on Ni's DOS, but with its atom type held fixed (e.g., pre-defined as Co).}
\label{fig:constrained-tasks}
\end{figure}

Here, we evaluate DOSMatGen's ability to achieve fine-grained control over atomic structure generation by applying targeted constraints to a specific subset of atoms within a single structure. Classifier-free guidance is utilized in all constrained generation. The three constrained generation tasks are 1) masked generation, 2) template guided generation and 3) fixed atom type generation. A visual representation of the three constrained generation tasks is shown in Fig. \ref{fig:constrained-tasks}. For demonstration, we employ the nickel DOS from Au$_3$Ni\footnote{Materials Project ID: \texttt{mp-976806}.} as the target DOS for conditional generation across all three tasks. Au$_3$Ni is chosen as a characteristic example of an alloy where the minority element, Ni in this case, exhibits ``free-atom like" (d-)states which are sharply localized in energy. Materials with sites exhibiting free-atom like states can have unique physical and chemical properties, including in applications towards catalysis \citep{greiner2018free, thirumalai2018investigating, fung2020electronic, rosen2023free}. Prior studies have shown that free-atom like states can induce strong chemisorption and facile bond activation, though materials which exhibit this property can be relatively rare. Therefore, using DOSMatGen to generate other instances of materials with free-atom like states serves as an illuminating test for the capability of this model to generalize well to rare cases in the training dataset. 


\subsubsection{Masked Generation}
\begin{figure}[h]
\centering
\begin{tikzpicture}
    \draw (0, 0) node[inner sep=0] {\includegraphics[width=\textwidth]{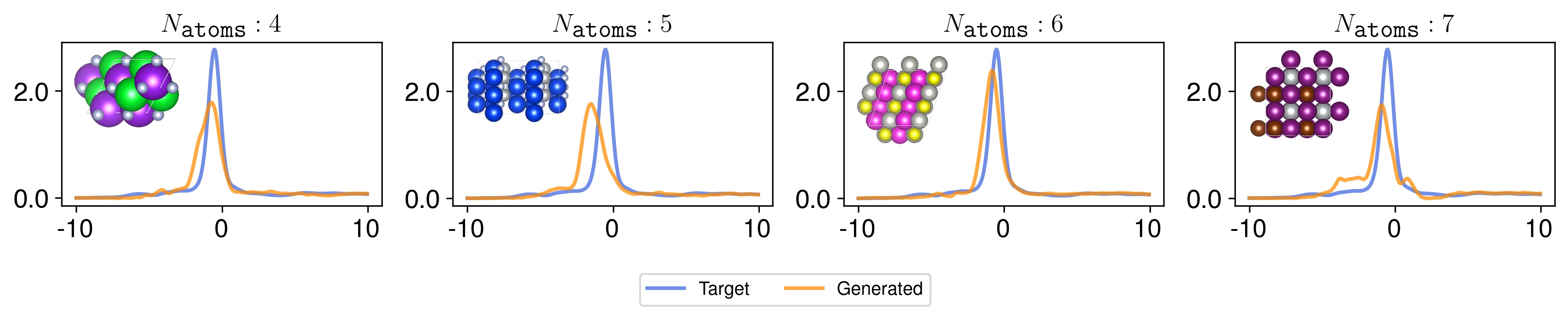}};
    \draw (-4.4, 0.9) node {\scriptsize{\color{red}Ni}KNSr};
    \draw (-0.7, 0.9) node {\scriptsize{\color{red}Ni}Ho$_2$N$_2$};
    \draw (2.9, 0.9) node {\scriptsize{\color{red}Ni}Cd$_2$Pd$_2$S};
    \draw (6.7, 0.9) node {\scriptsize{\color{red}Ni}Br$_2$I$_3$Mn};
\end{tikzpicture}
\caption{Density of states (DOS) for generated materials with $N_{\text{atoms}} \in \{4,5,6,7\}$. The orange curve represents the DOS of the first atom in each generated material, conditioned on Ni's DOS from Au$_3$Ni (shown in blue). The element type of the first atom is indicated in red. Visualizations of all structures are of a $2\times 2 \times 2$ supercell.}
\label{fig:masked-generation}
\end{figure}
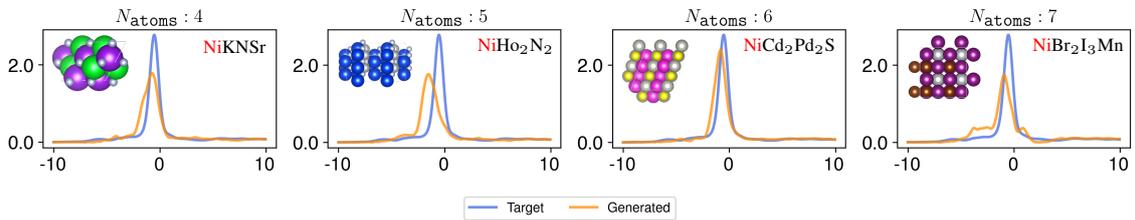

In the masked generation task, we seek to match the target DOS for an arbitrary subset of atoms, while allowing the DOS of the unconditioned atoms to vary freely, thereby enabling the generation of diverse structures that still meet the specified constraints. Effectively, we are generating new local atomic environments which yield the same desired DOS, which still come together to form a stable crystal structure. Constraining only a subset of atoms rather than the entire material also allows for greater diversity in the generated structures, without compromising on the overall goal of finding materials with free-atom like states. We begin the generation procedure at starting timestep $T=1000$, where $\mathbf{A}_T$, $\mathbf{X}_T$ and $\mathbf{L}_T$ are sampled from their respective prior distributions. In the masked generation case, we constrain the number of atoms to the set $N_{\text{atoms}} \in \{4,5,6,7\}$, though this can be readily changed as needed. For each value of $N_{\text{atoms}}$, we condition the first atom using Ni's DOS from Au$_3$Ni, while leaving the remaining atoms unconditioned. We then generate 50 candidate structures for each $N_{\text{atoms}}$ and select a representative structure to visualize its first atom's DOS with the ground truth in Fig. \ref{fig:masked-generation}.

As shown in Fig. \ref{fig:masked-generation}, the ground truth DOS of the nickel atom in Au$_3$Ni features a centralized sharp peak. Similarly, the four selected generated materials, each with a different $N_{\text{atoms}}$, also exhibit a centralized sharp peak near the ground truth. 
While the first atom in each generated material is explicitly conditioned on the reference Ni DOS, the other atoms are free to adopt different element types and locations. Consequently, the majority of the generated structures end up with Ni for the conditioned atom (even though it is not explicitly specified), while the other atoms in the unit cell end up with many diverse compositions. This test therefore successfully demonstrates the ability of DOSMatgen to recreate the Ni electronic structure within Au$_3$Ni in novel materials are dramatically different from the original reference material.



\subsubsection{Template Guided Generation}
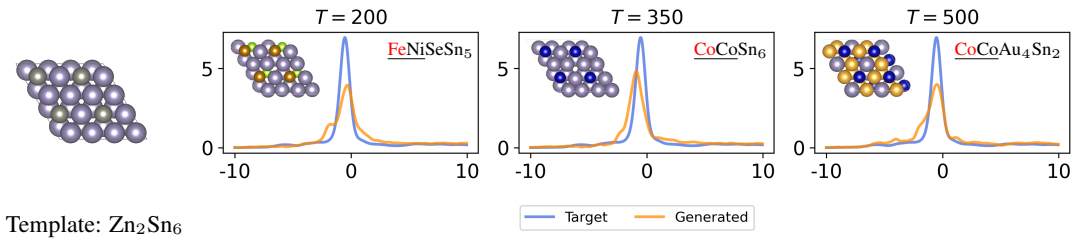
\begin{figure}[h]
     \centering
     {
     \captionsetup[subfigure]{labelformat=empty}
     \begin{subfigure}[b]{0.2\textwidth}
         \centering
         \raisebox{2.5mm}{ 
             \includegraphics[width=\textwidth]{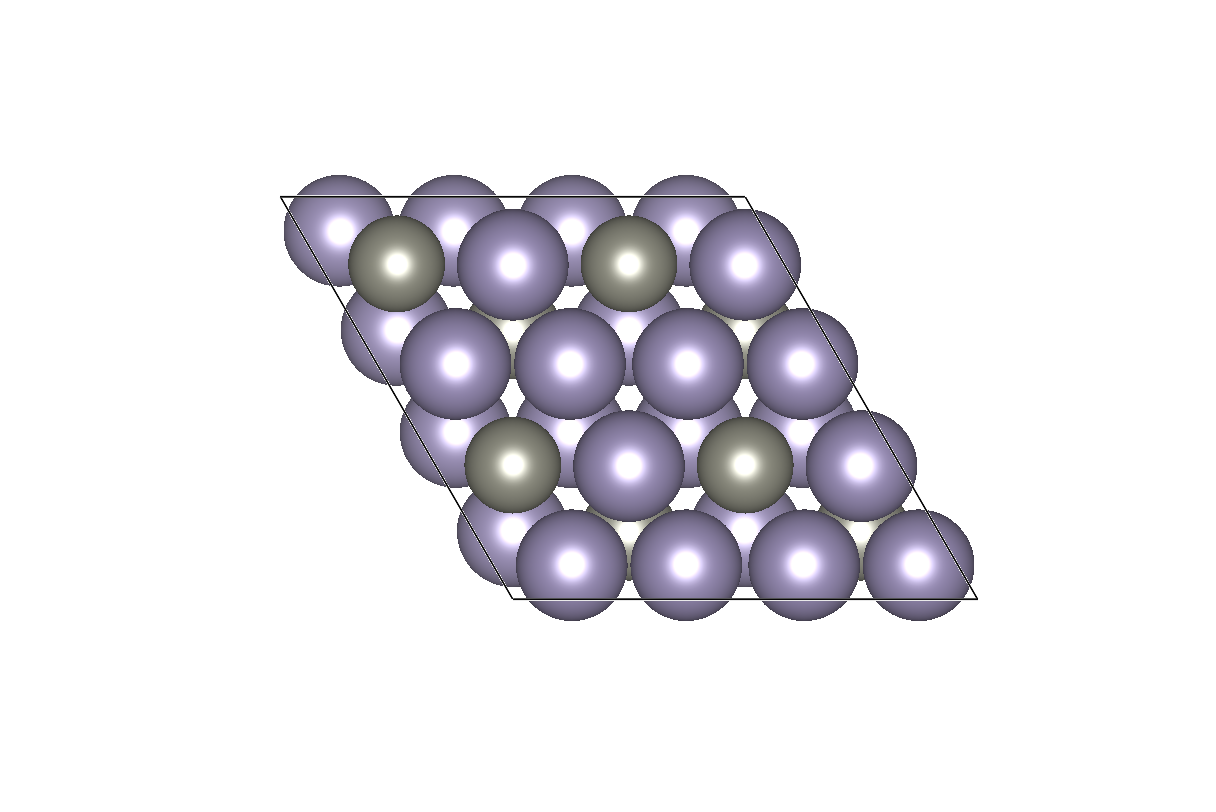}
         }
         \caption{$\qquad$Template: Zn$_2$Sn$_6$}
         \label{fig:template_zn2sn6}
     \end{subfigure}
     \begin{subfigure}[b]{0.78\textwidth}
         \begin{tikzpicture}
             \draw (0, 0) node[inner sep=0] {\includegraphics[width=\textwidth]{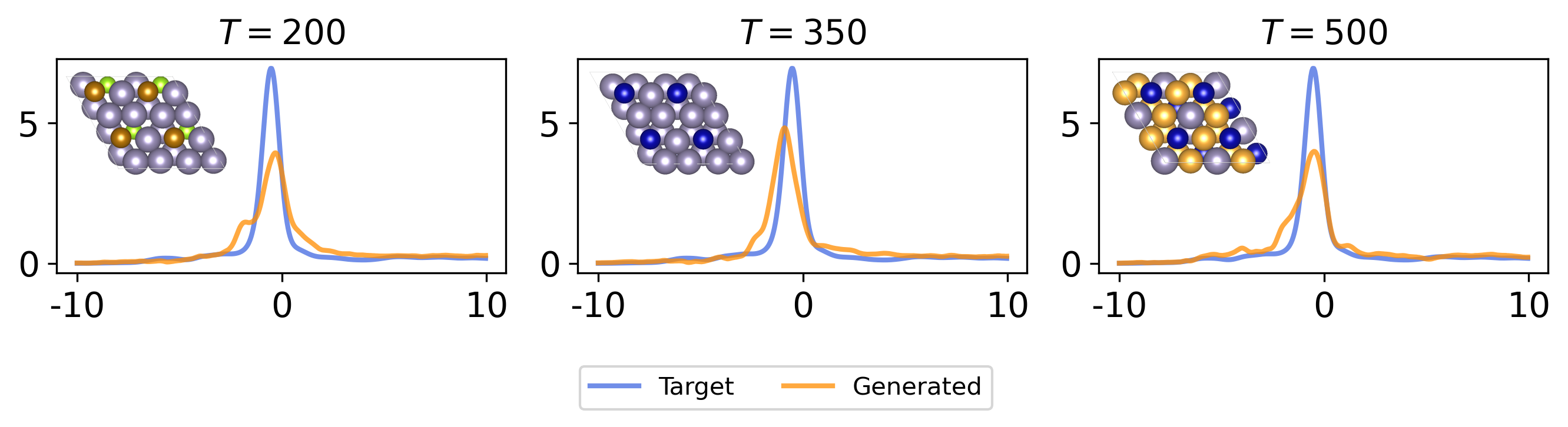}};
             \draw (-2.7, 0.9) node {\scriptsize{\underline{{\color{red}Fe}Ni}SeSn$_5$}};
             \draw (1.25, 0.9) node {\scriptsize{\underline{{\color{red}Co}Co}Sn$_6$}};
             \draw (4.9, 0.9) node {\scriptsize{\underline{{\color{red}Co}Co}Au$_4$Sn$_2$}};
         \end{tikzpicture}
     \end{subfigure}
     }
     \caption{Density of states (DOS) for generated materials from starting timestep $T \in \{200, 350, 500\}$. The template material Zn$_2$Sn$_6$ is shown on the far left. The orange curve represents the DOS of the first atom in each generated material, conditioned on Ni's DOS from Au$_3$Ni (shown in blue). The conditioned atoms are underlined, with the first atom highlighted in red to indicate that the plotted DOS corresponds to it. Visualizations of all structures are of a $2\times 2 \times 2$ supercell.}
     \label{fig:diff-start-mat}
\end{figure}

In this task, we investigate the use of an existing material to serve as a template for the diffusion process. Instead of sampling $\mathbf{A}_T$, $\mathbf{X}_T$, and $\mathbf{L}_T$ from their prior distributions at $T=1000$---representing the generation from complete noise---we initiate the diffusion process at $T<1000$. This approach involves adding noise to a specified template material at the chosen timestep. Similar to the masked generation task, only a subset of the atoms of the template material is conditioned on a target DOS, while the DOS of the unconditioned atoms can vary freely. This type of task can be useful when performing inverse design within specific materials classes such as perovskites or two-dimensional materials. For illustration, we use Zn$_2$Sn$_6$\footnote{Materials Project ID: \texttt{mp-971919}.} as the template material and initialize the diffusion process at 
$T\in\{200,350,500\}$. In this task, the Zn atoms in Zn$_2$Sn$_6$ are conditioned to match Ni's DOS. We generate 50 candidate structures for each $T$ and select a representative structure to visualize its conditioned atoms' DOS with the ground truth in Fig. \ref{fig:diff-start-mat}.

From Fig. \ref{fig:diff-start-mat}, we can observe that the DOS of the first atom in the generated materials closely aligns with the target DOS, but still maintains the 3D structure of the original template. Due to the diffusion process starting from a template material \textit{noisified} at $T<1000$, the final composition exhibits a strong bias towards the original Sn, though at higher $T$, the composition moves farther away from Sn to become other elements. Interestingly, the original Zn atoms, which are conditioned on Ni's DOS, are replaced by other element types such as Fe and Co rather than Ni. This suggests that DOSMatGen can propose chemically viable substitutions for the conditioned atoms by considering the overall composition, even when the target DOS is directly associated with that of Ni.

\subsubsection{Fixed Atom Type Generation}
\begin{figure}[h]
\centering
\begin{tikzpicture}
    \draw (0, 0) node[inner sep=0] {\includegraphics[width=\textwidth]{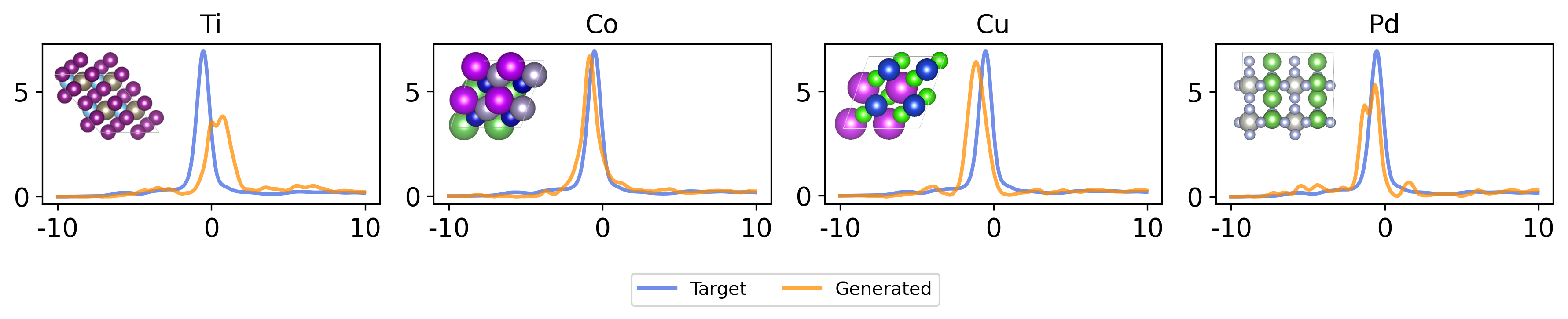}};
    \draw (-4.33, 0.9) node {\scriptsize{{\color{red}Ti}I$_3$Tl}};
    \draw (-0.8, 0.9) node {\scriptsize{{\color{red}Co}LaSnGd}};
    \draw (3.1, 0.9) node {\scriptsize{{\color{red}Cu}Cl$_2$Bi}};
    \draw (6.83, 0.9) node {\scriptsize{{\color{red}Pd}F$_5$As$_2$}};
\end{tikzpicture}
\caption{Density of states (DOS) for generated materials with the first atom constrained to Ti, Co, Cu or Pd. The orange curve represents the DOS of the first atom in each generated material, conditioned on Ni's DOS from Au$_3$Ni (shown in blue). Visualizations of the corresponding structures are of a $2\times 2 \times 2$ supercell.}
\label{fig:fix-atom-type}
\end{figure}
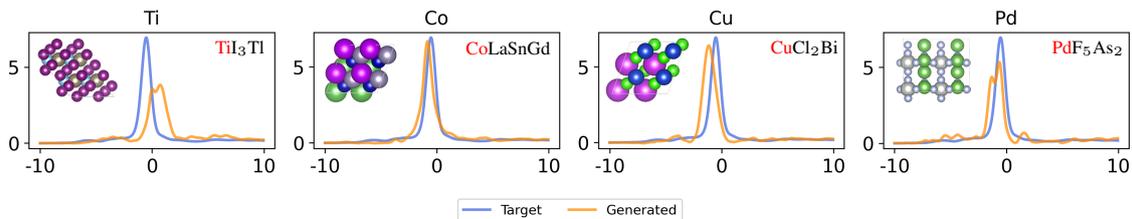

In this task, we explore the feasibility of fixing the first atom's element type to an arbitrary specified element. Similar to the masked generation task, the diffusion process begins at $T=1000$, meaning the materials are generated entirely from noise. The key difference in this task is that the first atom is constrained to a predefined element type, whereas in the masked generation task, the model can freely diffuse the first atom into any element type. The first atom in the generated structure is still conditioned on the DOS of Ni from Au$_3$Ni. In our experiments, we define a constraint on the element type to be Ti, Co, Cu, or Pd.
Co and Cu are selected because their proximity to to Ni in the same period, and similarly Pd is chosen for being in the same group as Ni. Meanwhile, Ti is included as an element which bears little chemical similarity to Ni, to test if the model can nevertheless generate an atomic environment to enable Ti to behave electronically like Ni. For each element type, we generate 50 candidate structures and select a representative structure to visualize its first atom's DOS with the ground truth in Fig. \ref{fig:fix-atom-type}.

Fig. \ref{fig:fix-atom-type} reveals that when the first atom is constrained to neighboring transition metals such as Co, Cu, and Pd, the resulting density of states (DOS) demonstrates remarkable alignment with the target DOS. This is to be expected due to the proximity of Pd, Co and Cu to Ni in the periodic table. 
Meanwhile, the DOS of the Ti-constrained case is not as similar to target DOS, though it still exhibits a peak near the ground truth peak location, suggesting that this is a harder task, but one which DOSMatGen is still able to follow the guidance to some extent. In addition, in all four generated structures shown in Fig. \ref{fig:fix-atom-type}, we observe high compositional diversity in the unconditioned atoms. These are exciting observations, as it suggests that DOSMatGen can provide viable elemental substitutions while maintaining a similar electronic structure. This concept can conceivably be expanded to tackle pressing materials design challenges such as finding earth-abundant and low-toxicity substitutions to existing functional materials such as noble metal catalysts or lead-based halide perovskites. 


\subsection{Post-processing Workflow}
The results provided so far provide illustrative examples of the capabilities of DOSMatGen for generating well-conditioned crystal structures under various constraints. Here, we show how DOSMatGen can be further incorporated into a more comprehensive materials discovery workflow to generate large quantities of high-quality materials. The workflow involves using DOSMatGen to first conditionally generate many candidate structures (i.e. 10,000 or more), followed by structural relaxation using a universal machine learning force field (MLFF) to yield local minimum structures. After relaxation, the formation energy per atom (in eV/atom) is calculated for all structures based on MLFF energies. Structures with formation energies (or optionally, energy below hull) below a desired threshold are then filtered. From this subset, the predicted DOS of the generated materials is compared against the ground truth DOS, and the best-matched structures are then selected for further evaluation with DFT calculations. These post-processing steps help ensure the generated materials will all be locally minimum, stable, and well-matched in terms of the DOS. 

We select the DOS of Ir from rutile IrO$_2$ as the target, which presents more a complex target DOS than the previous example. IrO$_2$ is widely considered to be a state-of-the-art electrocatalyst for oxygen evolution \citep{seitz2016highly,song2020review, wei2019recommended}. Designing new materials which exhibit the electronic structure of Ir could help lead to similarly well-performing catalysts. Here we first generate 10,000 candidate structures, with number of atoms in the unit cell within the range $N_{\text{atoms}} \in \{4, 5, 6, 7, 8\}$. For each value of $N_{\text{atoms}}$, 2,000 structures are generated, ensuring an even distribution across the possible values. We then use \texttt{Orb-v2} \citep{neumann2024orb} as the MLFF for structure relaxation and for calculating the formation energy per atom for all structures. Structures with formation energies $\le -1.5$ eV/atom are then selected. 

\begin{figure}[h]
\centering
\begin{tikzpicture}
    \draw (0, 0) node[inner sep=0] {\includegraphics[width=0.65\textwidth]{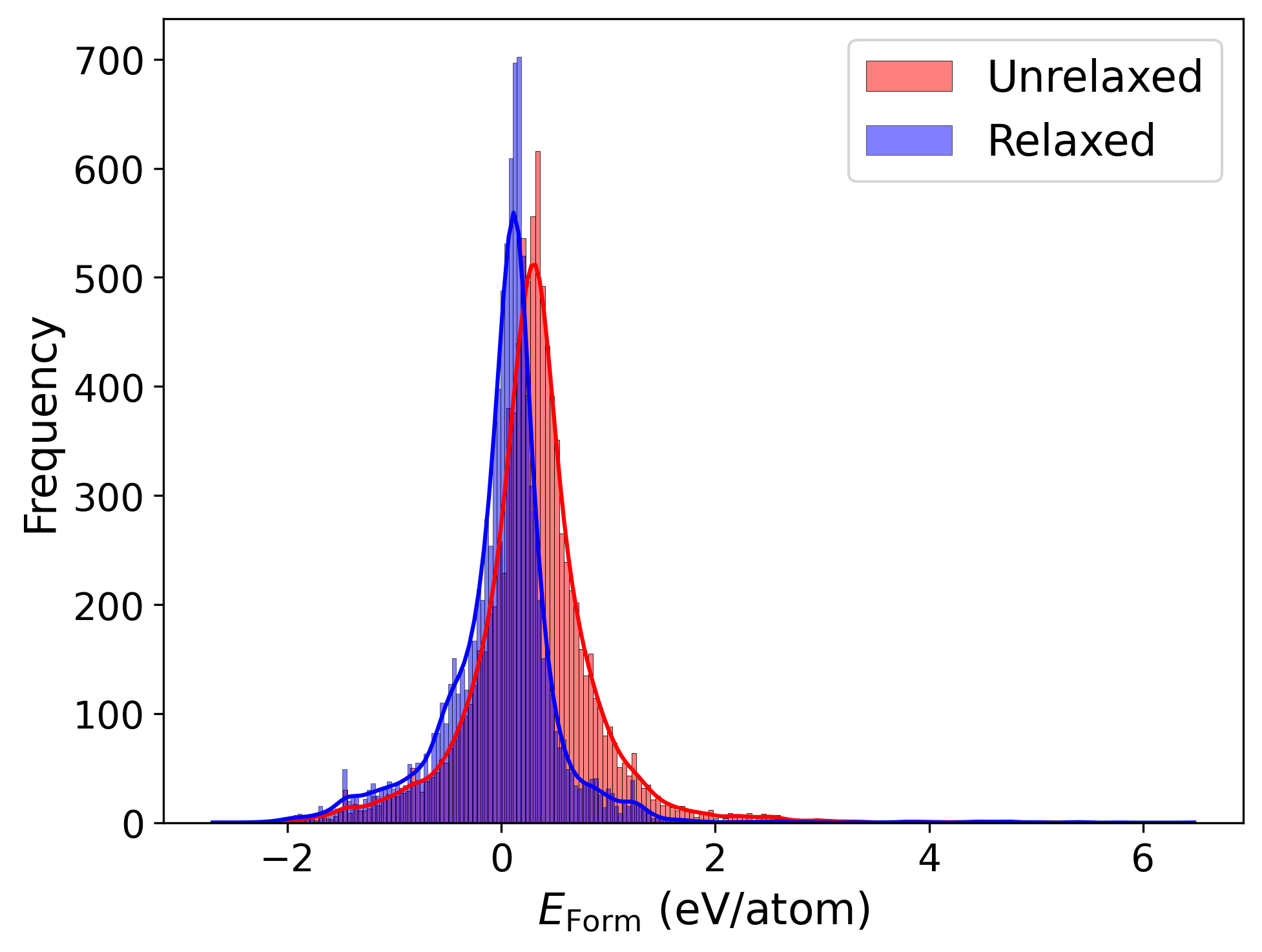}};
\end{tikzpicture}
\caption{Distribution of formation energies per atom (eV/atom) for 10,000 generated structures before (unrelaxed) and after (relaxed) structural optimization using \texttt{Orb-v2}. The histogram shows the frequency of structures across different formation energies.}
\label{fig:iro2-distribution}
\end{figure}

Fig. \ref{fig:iro2-distribution} shows the distribution of formation energies per atom for the 10,000 unrelaxed and relaxed structures. As expected, the distribution for the relaxed structures is shifted to the left compared to the unrelaxed structures, demonstrating that structural relaxation with an MLFF effectively lowers the formation energies. Notably, both distributions closely resemble Gaussian distributions centered around zero, with significant overlap between them. This indicates that DOSMatGen can conditionally generate relatively stable crystal structures even without structural optimization, and the the MLFF is mainly used for providing further refinement.
 
\begin{figure}[h]
\centering
\begin{tikzpicture}
    \draw (0, 0) node[inner sep=0] {\includegraphics[width=\textwidth]{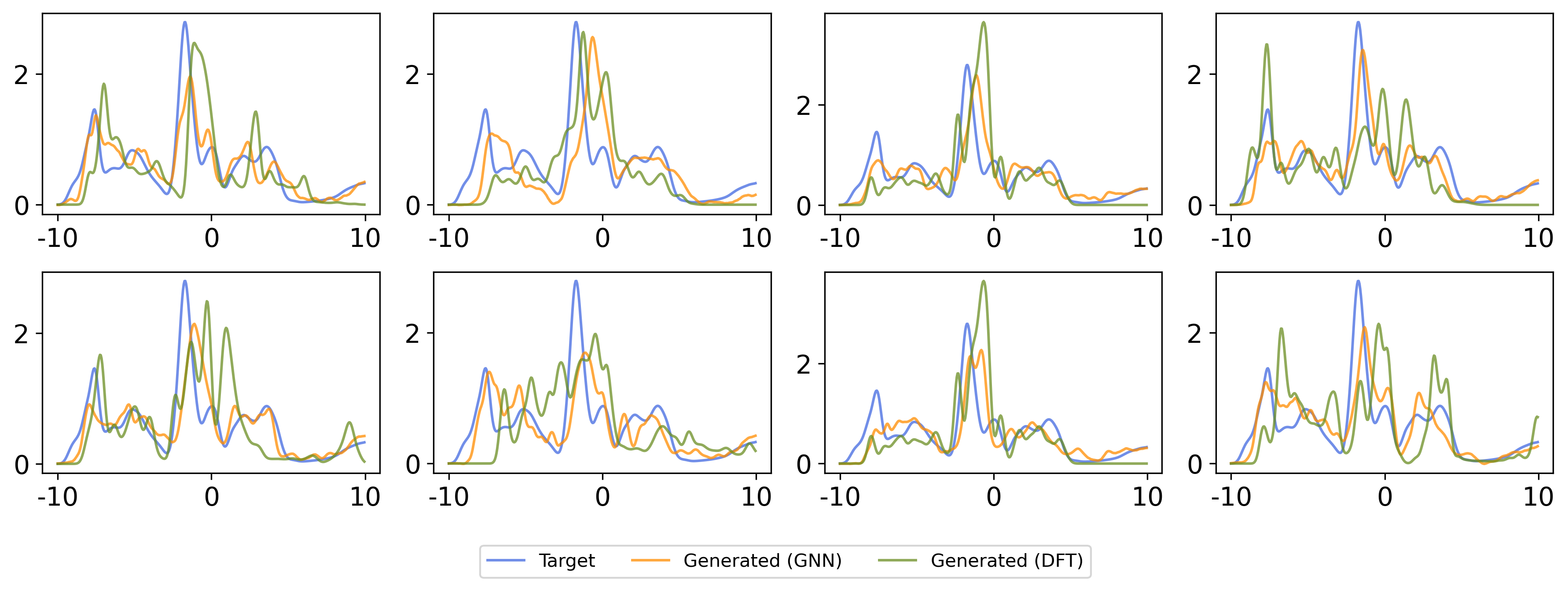}};
    \draw (-4.3, 2.4) node {\scriptsize{\color{red}Ir}HfO$_4$};
    \draw (-0.7, 2.4) node {\scriptsize{\color{red}Ru}SmO$_2$};
    \draw (3.1, 2.4) node {\scriptsize{\color{red}Ir}Ta$_2$O$_5$};
    \draw (6.95, 2.4) node {\scriptsize{\color{red}Ir}TiO$_4$};
    \draw (-4.3, -0.1) node {\scriptsize{\color{red}Ir}TaO$_4$};
    \draw (-0.7, -0.1) node {\scriptsize{\color{red}Ir}HoO$_2$};
    \draw (3.1, -0.1) node {\scriptsize{\color{red}Ir}Ti$_2$SO$_4$};
    \draw (6.95, -0.1) node {\scriptsize{\color{red}Ir}AlO$_4$};
\end{tikzpicture}
\caption{Density of states (DOS) for generated materials with the first atom conditioned on the DOS of Ir from IrO$_2$. The blue curve represents the ground truth target DOS. The orange curve represents the DOS of the first atom in each generated material, predicted using the forward GNN model. The green curve represents the DOS of the first atom computed via DFT. The first atom's element type is highlighted in red in the composition.}
\label{fig:iro2-screening}
\end{figure}
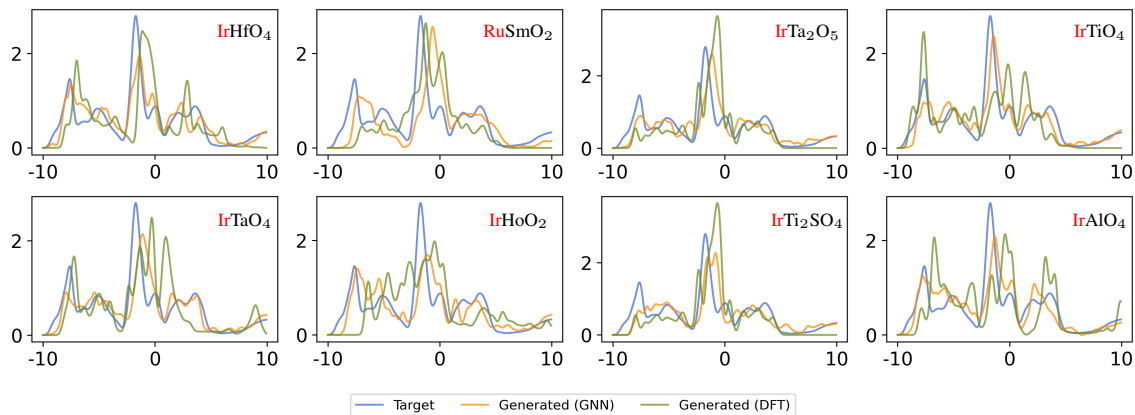

Applying a formation energy threshold of $\leq -1.5$ eV/atom yields a total of 108 candidate structures, from which 8 are selected for further DFT validation based on closeness to the target DOS. Fig. \ref{fig:iro2-screening} compares the target DOS with the generated DOS, obtained using either the surrogate GNN model or computed from DFT. While the 8 selected structures do not perfectly reproduce all features of the target DOS, they generally show strong resemblance, capturing key peaks and overall trends. This remains true under DFT validation as well. Visualizations of the eight selected structures are provided in Appendix \ref{sec:appendix-c}. Here, we note that the final materials all tend towards containing Ir in some rutile or rutile-like configuration. It is an unsurprising result, given the target DOS is derived from Ir in rutile IrO$_2$, though it fails to yield the desired earth-abundant alternative materials. This may be attributed to the fact that matching the DOS of Ir exactly may be too strong a constraint leading to low diversity, whereas it may only be necessary to match certain regions in the DOS (i.e. near the Fermi level). At the same time, it may also suggest the intrinsic difficulty of this particular problem, which we leave for future investigation. 

\section{Conclusion}\label{sec:conclusion}
In this work, we introduced DOSMatGen, an E(3)-equivariant joint diffusion framework for the conditional generation of crystalline materials guided by quantities from electronic structure such as the DOS. DOSMatGen excels in generating compositionally diverse and stable structures while maintaining strong alignment with target properties. The model demonstrates remarkable versatility across constrained generation tasks, including masked generation, initialization from template materials, and atom-type constraints, enabling precise and fine-grained control over the generated structures. Additionally, through a large-scale screening approach, DOSMatGen effectively identifies stable candidate materials with complex electronic properties such as those of IrO$_2$. The ability to balance structural validity, property alignment, and compositional diversity positions DOSMatGen as a valuable tool for a wide range of materials discovery tasks, guided by the physically important objective of the electronic structure. Finally, it is worth noting that DOSMatGen is property-agnostic and can be readily adapted to other electronic and even spectroscopic properties, including 1D X-ray diffraction (XRD) spectra, angle-resolved photoemission spectra (ARPES), X-ray absorption spectra (XAS) among others, as well as 2D images such as scanning transmission electron microscopy (STEM) images. 


\section{Data Availability}
The authors declare that all data, materials, and code supporting the results reported in this study are available at the following repository: \url{https://github.com/Fung-Lab/DOSMatGen}.

\section{Acknowledgements}
This research used resources of the National Energy Research Scientific Computing Center (NERSC), a U.S. Department of Energy Office of Science User Facility located at Lawrence Berkeley National Laboratory, operated under Contract No. DE-AC02-05CH11231 using NERSC award BES-ERCAP0022842. Part of this research is sponsored by the INTERSECT Initiative as part of the Laboratory Directed
Research and Development Program of Oak Ridge National Laboratory, managed by UT-
Battelle, LLC, for the U.S. Department of Energy under contract DE-AC05-00OR22725, via
the QCAD project (PI:P.Ganesh). Part of this research used resources of the National Energy Research Scientific Computing Center, a DOE Office of Science User Facility
supported by the Office of Science of the U.S. Department of Energy under Contract No. DE-AC02-05CH11231 using NERSC award BES-ERCAP0031261, as part of a user project at the Center for Nanophase Materials Sciences (CNMS), which is a US Department of Energy, Office of Science User Facility at Oak Ridge National Laboratory.




\clearpage
\bibliography{refs}
\bibliographystyle{naturemag}
\clearpage

\appendix
\counterwithin{figure}{section}
\counterwithin{table}{section}
\counterwithin*{equation}{section}
\renewcommand\theequation{\thesection\arabic{equation}}

\begin{titlepage}
  \centering
  \LARGE \textsc{Supplementary Information} \par
  \let\endtitlepage\relax
\end{titlepage}

\section{Additional Details on Joint Equivariant Diffusion}\label{sec:appendix-a}
\textbf{Diffusion on Lattice} $\mathbf{L}$. Inspired by DiffCSP \citep{jiao2024crystal}, the forward noising process that progressively adds noise to $\mathbf{L}$ is defined as
\begin{equation}
    q\left(\mathbf{L}_t|\mathbf{L}_0\right) = \mathcal{N}\left(
        \mathbf{L}_t | \sqrt{\bar{a}_t}\mathbf{L}_0, \left( 1-\bar{a}_t \right)\mathbf{I}
    \right),
\end{equation}
where $\mathbf{L}_t$ is the noisy version of $\mathbf{L}_0$ at timestep $t$, and $\bar{a}_t = \Pi^t_{s}\left( 1-\beta_t \right)$ determines how much of the original data, $\mathbf{L}_0$, is preserved at $t$ given $\beta_t\in (0,1)$ \citep{nichol2021improved}. During the diffusion process, we start with $\mathbf{L}_T\sim \mathcal{N}\left(0,\mathbf{I}\right)$, and use the trained denoising neural network to obtain $\mathbf{L}_{t-1}$ from $\mathbf{L}_t$:
\begin{align}
    p\left( \mathbf{L}_{t-1} | S_t \right) &= \mathcal{N}\left(
        \mathbf{L}_{t-1} | \mu \left(S_t\right), \sigma^2\left(S_t\right)\mathbf{I}
    \right), \\
    \mu \left(S_t\right) &= \frac{1}{\sqrt{\alpha_t}} \left(
        \mathbf{L}_t - \frac{\beta_t}{\sqrt{1-\bar{a}_t}}\hat{\boldsymbol{\epsilon}}_{\mathbf{L}}\left(S_t, t\right)
    \right),\label{eq:A-lattice-mu} \\
    \sigma^2\left(S_t\right) &= \beta_t \left( \frac{1-\bar{a}_{t-1}}{1-\bar{a}_t} \right). \label{eq:A-lattice-sigma}
\end{align}
Here, $\hat{\boldsymbol{\epsilon}}_{\mathbf{L}}$ is predicted by the denoising network given the intermediate noisy structure $S_t = \left(\mathbf{A}_t, \mathbf{X}_t, \mathbf{L}_t\right)$.

\textbf{Diffusion on Fractional Coordinates} $\mathbf{X}_f$. Following DiffCSP, we use wrapped normal distribution as the prior for $\mathbf{X}_f$ with the following forward noising process:
\begin{equation}
    q\left(\mathbf{X}_t|\mathbf{X}_0\right) \varpropto \sum_{\mathbf{Z}\in \mathbb{Z}^{3\times N}} \exp\left(
        -\frac{\|\mathbf{X}_t - \mathbf{X}_0 + \mathbf{Z}\|^2}
        {2\sigma^2_t}
    \right),
\end{equation}
which ensures the probability distribution over $[z,z+1)^{3\times N}$ for any integer $z$ is the same. Here, $\sigma_t = \sigma_1\left(\frac{\sigma_T}{\sigma_1}\right)^{\frac{t-1}{T-1}}$ for $t>0$ and $\sigma_0 = 0$. During the diffusion process, we start with $\mathbf{X}_T \sim \mathcal{U}(0,1)$, and use the trained denoising neural network to obtain $\mathbf{X}_{t-1}$ from $\mathbf{X}_t$ using the denoising term $\hat{\boldsymbol{\epsilon}}_{\mathbf{X}}$. 

\textbf{Diffusion on Atom Types} $\mathbf{A}$. We consider the atomic numbers of a crystal structure $S$ to be a continuous variable in real space $\mathbb{R}^{h\times N}$, where $h$ is the latent dimension of the atoms and $N$ is the total number of atoms. The atom types are initially one-hot encoded. Similar to the diffusion process for $\mathbf{L}$, the forward noising process and the backward diffusing process are
\begin{align}
q\left(\mathbf{A}_t|\mathbf{A}_0\right) &= \mathcal{N}\left(
        \mathbf{A}_t | \sqrt{\bar{a}_t}\mathbf{A}_0, \left( 1-\bar{a}_t \right)\mathbf{I}
    \right), \\
p\left(\mathbf{A}_{t-1}|S_t\right) &= \mathcal{N}\left(
    \mathbf{A}_{t-1} | \mu_{\mathbf{A}} \left(S_t\right), \sigma^2_{\mathbf{A}}\left(S_t\right)\mathbf{I}
\right),
\end{align}
where $\mu_{\mathbf{A}}$ and $\sigma^2_{\mathbf{A}}$ are defined in a similar way as Eqs. \ref{eq:A-lattice-mu} and \ref{eq:A-lattice-sigma}.

\textbf{Joint Diffusion Process}.
The training objectives for $\mathbf{A}$, $\mathbf{X}$ and $\mathbf{L}$ are:
\begin{align}
\mathcal{L}_{\mathbf{L}}&=\mathbb{E}_{\boldsymbol{\epsilon}_{\mathbf{L}} \sim \mathcal{N}(0, I), t \sim \mathcal{U}(1, T)}\left[\left\|\boldsymbol{\epsilon}_{\mathbf{L}}-\hat{\epsilon}_{\mathbf{L}}\left(\mathcal{S}_t, t\right)\right\|_2^2\right], \\
\mathcal{L}_{
\mathbf{X}}&=\mathbb{E}_{
\mathbf{X}_t \sim q\left(
\mathbf{X}_t \mid 
\mathbf{X}_0\right), t \sim \mathcal{U}(1, T)}\left[\lambda_t\left\|\nabla_{
\mathbf{X}_t} \log q\left(
\mathbf{X}_t \mid 
\mathbf{X}_0\right)-\hat{\boldsymbol{\epsilon}}_{
\mathbf{X}}\left(\mathcal{S}_t, t\right)\right\|_2^2\right],\\
\mathcal{L}_{\mathbf{A}} &= \mathbb{E}_{\boldsymbol{\epsilon}_{\mathbf{A}} \sim \mathcal{N}(0, \mathbf{I}), \, t \sim \mathcal{U}(1, T)} 
\left[ \left\| \boldsymbol{\epsilon}_{\mathbf{A}} - \hat{\boldsymbol{\epsilon}}_{\mathbf{A}}(\mathbf{S}_t, t) \right\|_2^2 \right],
\end{align}
where $\lambda_t = \mathbb{E}_{\mathbf{X}_t}^{-1} \left[ \left\| \nabla_{\mathbf{X}_t} \log q(\mathbf{X}_t \mid \mathbf{X}_0) \right\|_2^2 \right]$.

The overall combined objective for training the denoising neural network is then
\begin{align}
    \mathcal{L}_S = \lambda_{\mathbf{A}}\mathcal{L}_{\mathbf{A}}+ \lambda_{\mathbf{X}}\mathcal{L}_{\mathbf{X}} + \lambda_{\mathbf{L}}\mathcal{L}_{\mathbf{L}},
\end{align}
as shown in Eq. \ref{eq:combined-training-obj}.
\clearpage
\section{Additional Distribution Plots of Formation Energies Per Atom}\label{sec:appendix-b}
\begin{figure}[h]
     \centering
     \begin{subfigure}[b]{0.485\textwidth}
         \centering
         \includegraphics[width=\textwidth]{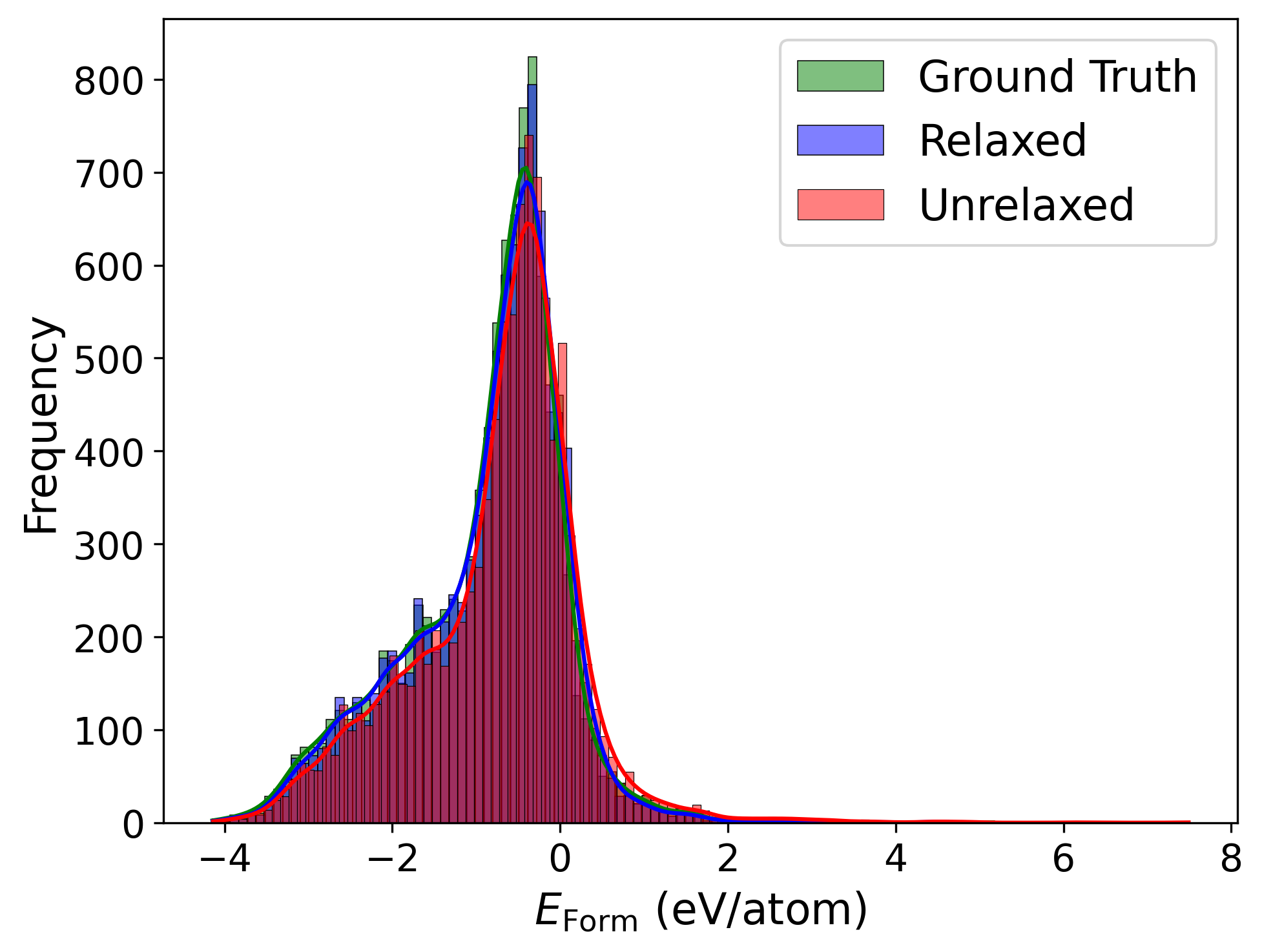}
         \caption{Classifier-based guidance at $T=200$}
         \label{fig:cg_distribution_200}
     \end{subfigure}
     \begin{subfigure}[b]{0.485\textwidth}
         \centering
         \includegraphics[width=\textwidth]{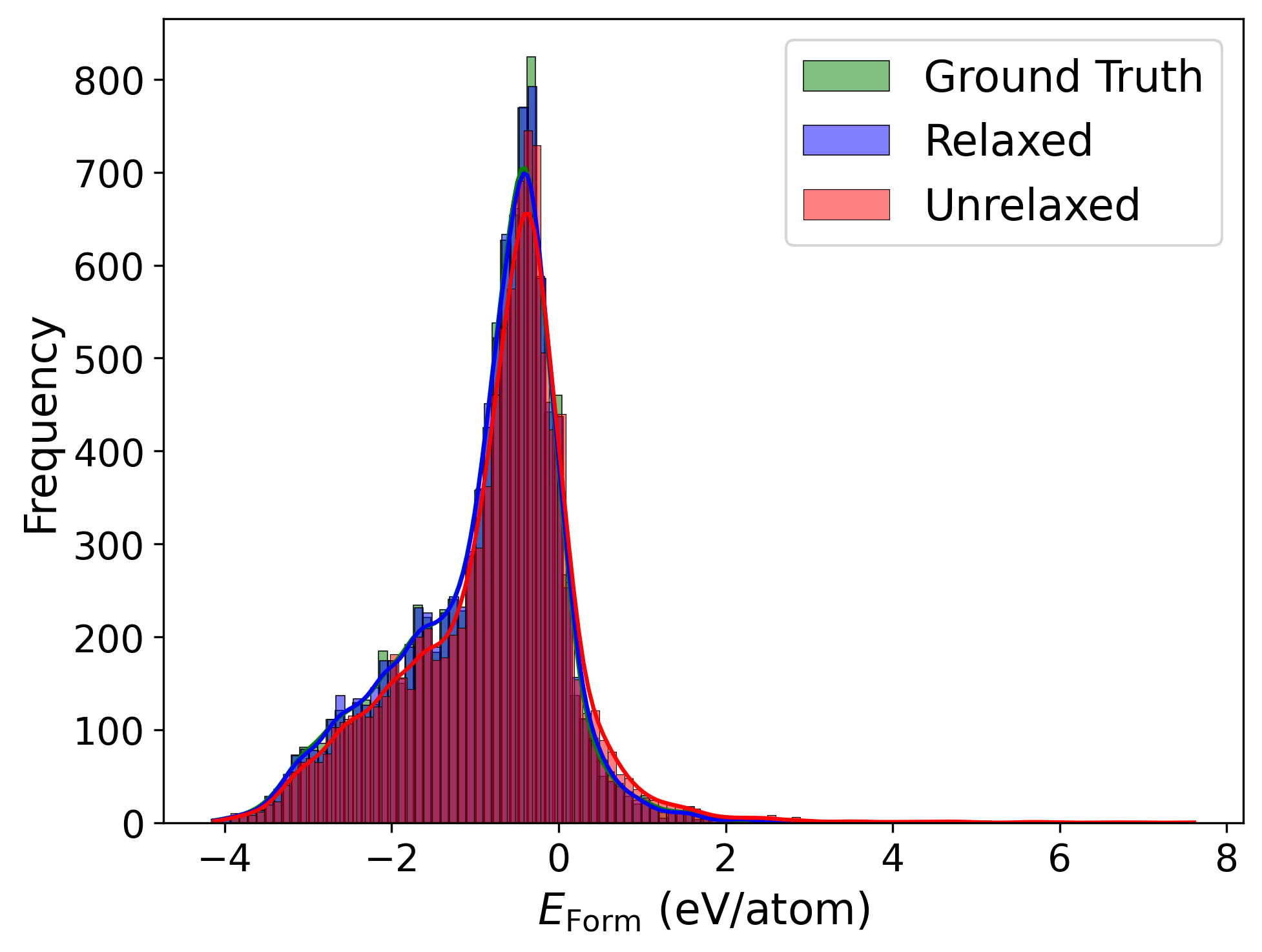}
         \caption{Classifier-free guidance at $T=200$}
         \label{fig:cfg_distribution_200}
     \end{subfigure}
     \begin{subfigure}[b]{0.485\textwidth}
         \centering
         \includegraphics[width=\textwidth]{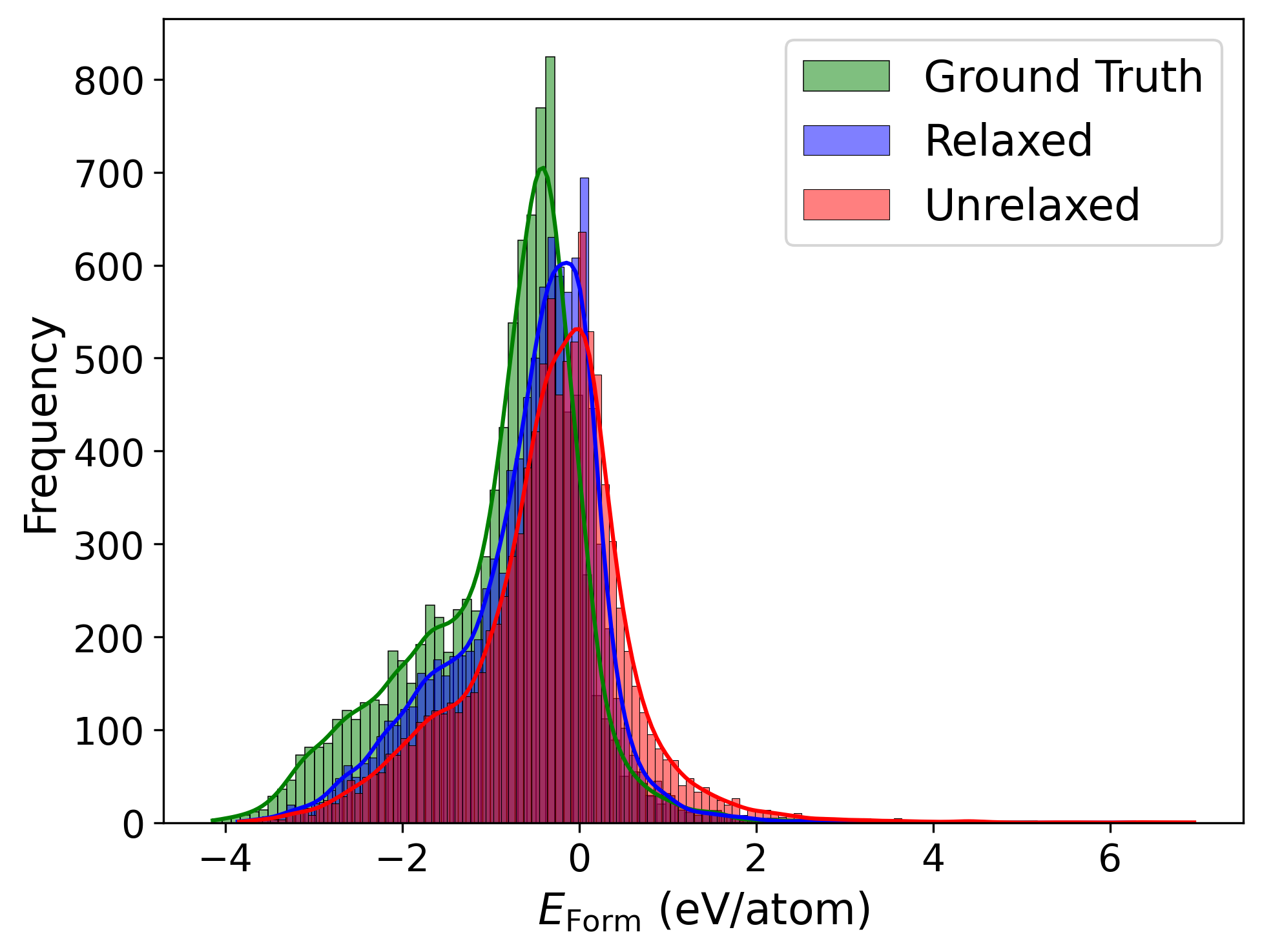}
         \caption{Classifier-based guidance at $T=500$}
         \label{fig:cg_distribution_500}
     \end{subfigure}
     \begin{subfigure}[b]{0.485\textwidth}
         \centering
         \includegraphics[width=\textwidth]{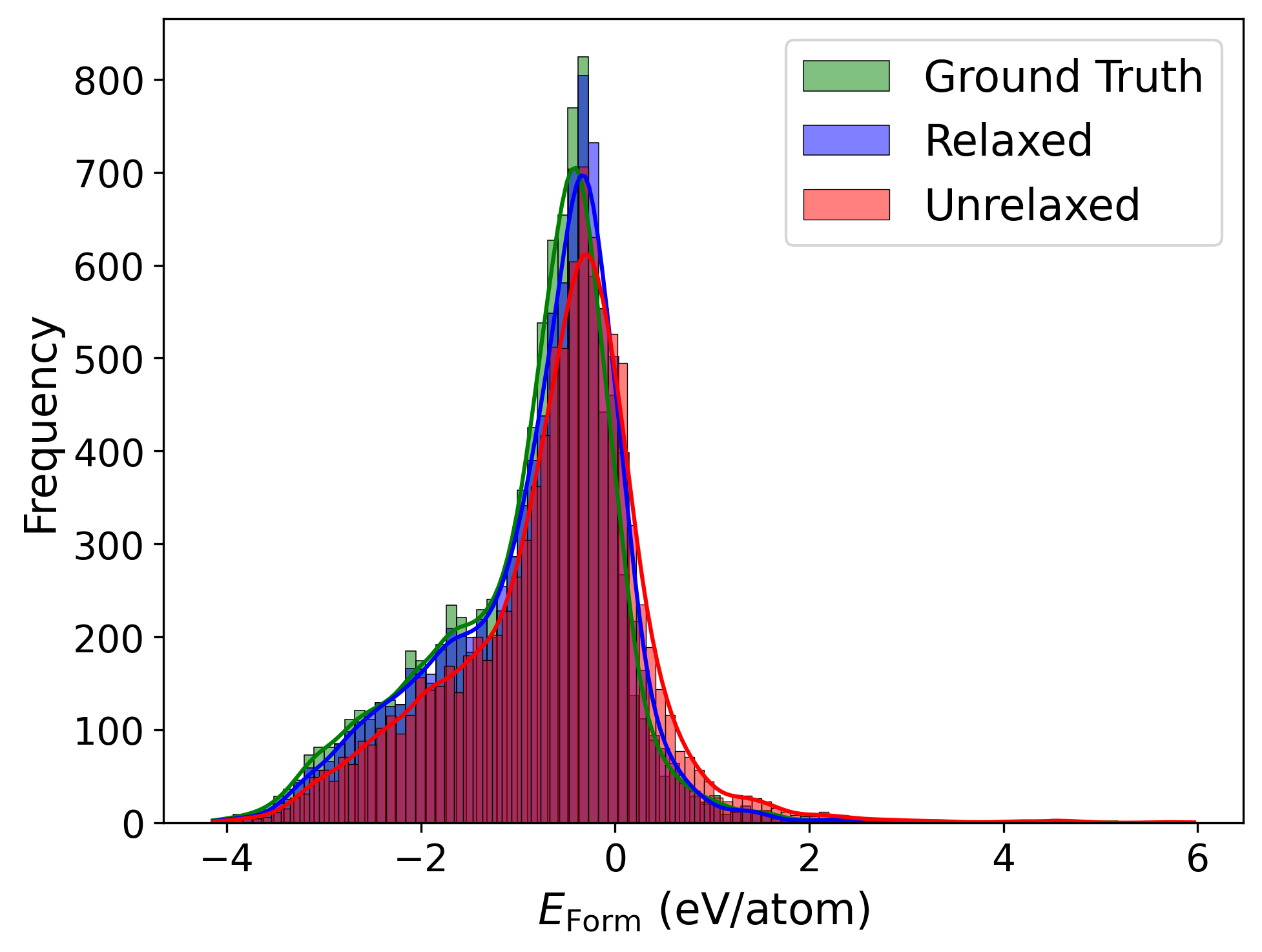}
         \caption{Classifier-free guidance at $T=500$}
         \label{fig:cfg_distribution_500}
     \end{subfigure}
        \caption{Distribution of formation energies per atom (eV/atom) for structures conditionally generated at $T\in\{200, 500\}$ using (a) classifier-based guidance and (b) classifier-free guidance. All generated structures are relaxed using the \texttt{Orb-v2} interatomic potential. The ground truth distribution corresponds to the formation energies of the original test set structures, with all energies predicted using \texttt{Orb-v2}.}
        \label{fig:testset_distributions_200_500}
\end{figure}

\clearpage
\section{Visualizations of Selected Structures from Post-processing Workflow}\label{sec:appendix-c}
The following visualizations correspond to the structures in Fig. \ref{fig:iro2-screening}.

\begin{figure}[h]
     \centering
     \begin{subfigure}[b]{0.2\textwidth}
        \includegraphics[width=\textwidth]{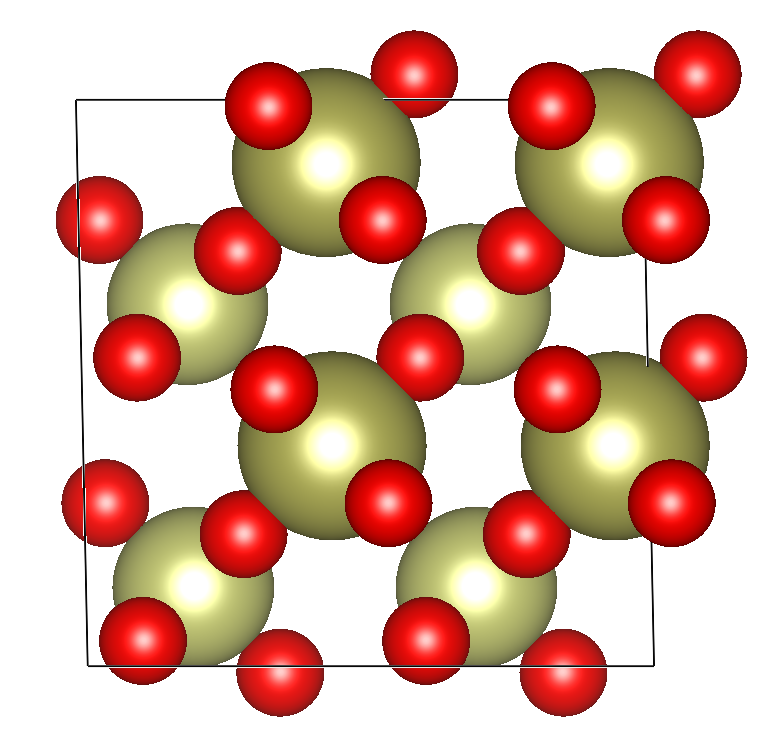}
         \caption{IrHfO$_4$}
         \label{fig:template_zn2sn6}
     \end{subfigure}
     \begin{subfigure}[b]{0.2\textwidth}
         \includegraphics[width=\textwidth]{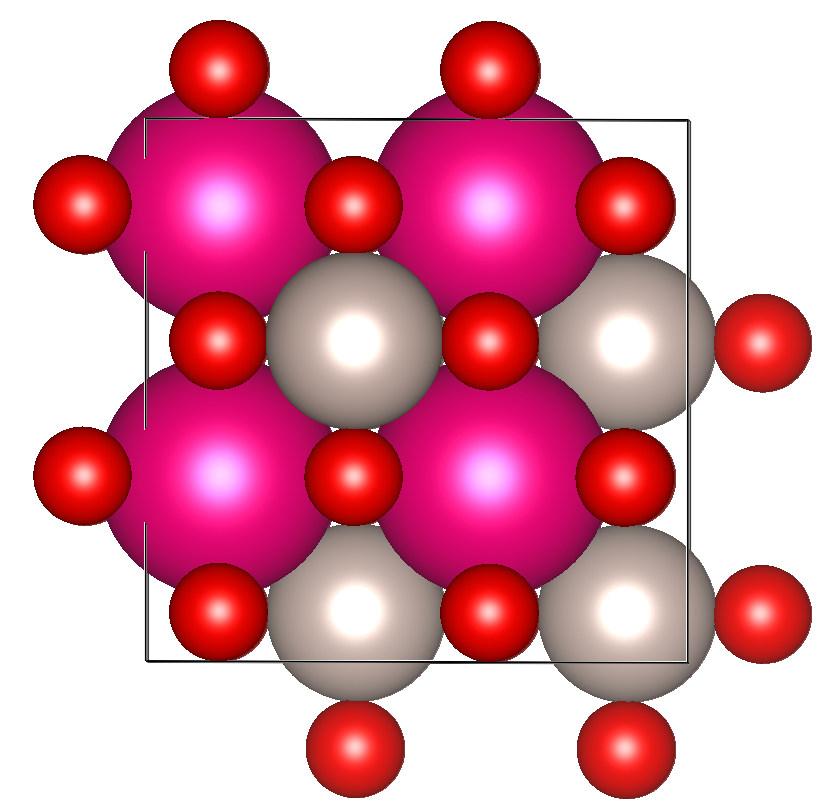}
         \caption{RuSmO$_2$}
     \end{subfigure}
     \begin{subfigure}[b]{0.2\textwidth}
        \includegraphics[width=\textwidth]{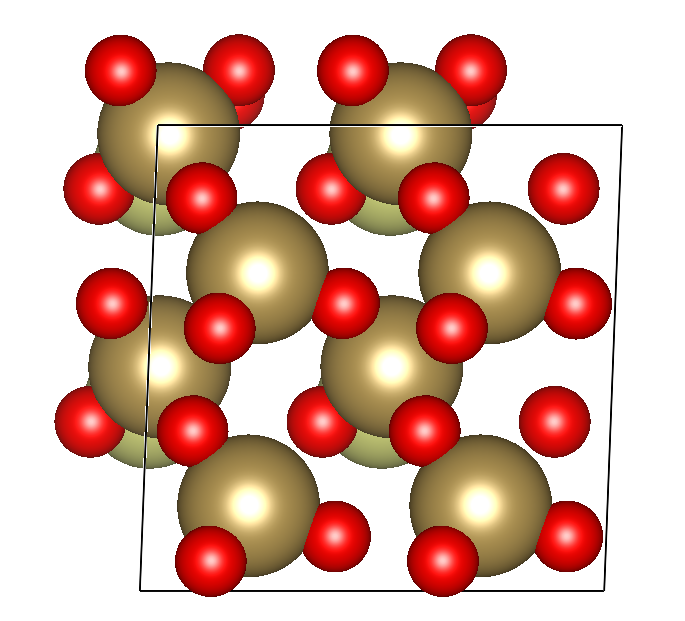}
         \caption{IrTa$_2$O$_5$}
         \label{fig:template_zn2sn6}
     \end{subfigure}
     \begin{subfigure}[b]{0.2\textwidth}
         \includegraphics[width=\textwidth]{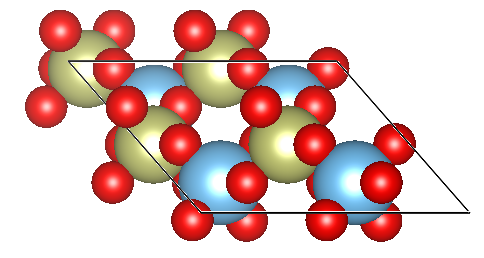}
         \caption{IrTiO$_4$}
     \end{subfigure}
     \begin{subfigure}[b]{0.2\textwidth}
        \includegraphics[width=\textwidth]{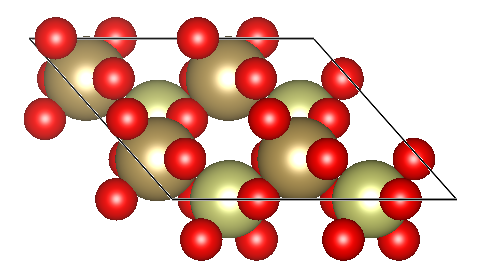}
         \caption{IrTaO$_4$}
         \label{fig:template_zn2sn6}
     \end{subfigure}
     \begin{subfigure}[b]{0.2\textwidth}
         \includegraphics[width=\textwidth]{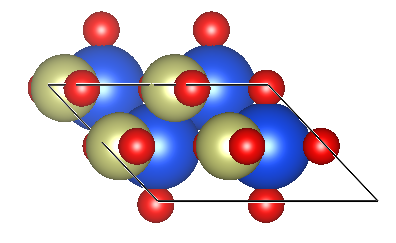}
         \caption{IrHoO$_2$}
     \end{subfigure}
     \begin{subfigure}[b]{0.2\textwidth}
        \includegraphics[width=\textwidth]{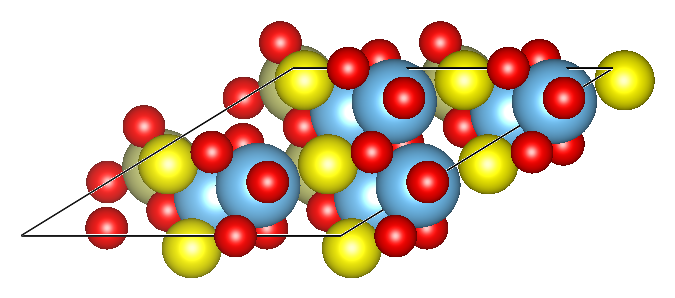}
         \caption{IrTi$_2$SO$_4$}
         \label{fig:template_zn2sn6}
     \end{subfigure}
     \begin{subfigure}[b]{0.2\textwidth}
         \includegraphics[width=\textwidth]{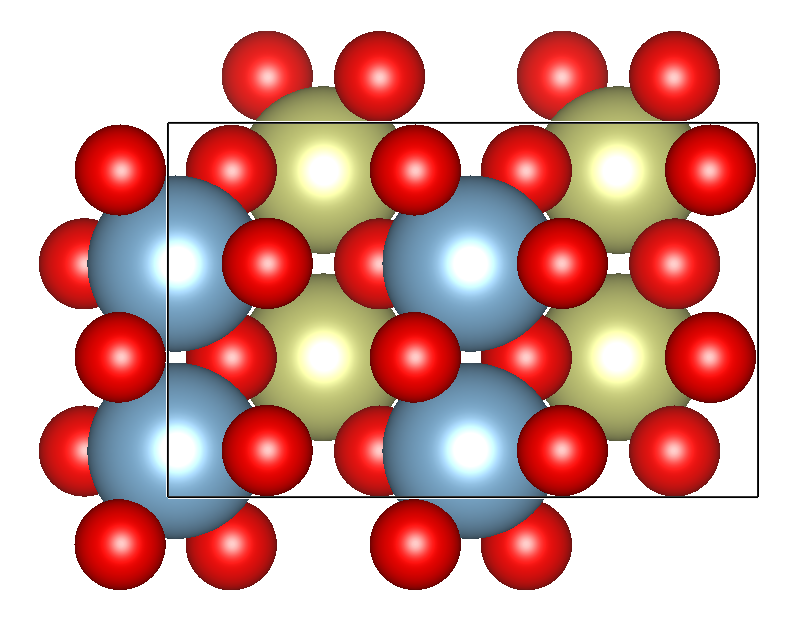}
         \caption{IrAlO$_4$}
     \end{subfigure}
     \caption{Visualizations of the crystal structures generated during the post-processing workflow (refer to Fig. \ref{fig:iro2-screening}), shown as $2\times 2 \times 2$ supercells.}
     \label{fig:iro2-screening-structures}
\end{figure}

\end{document}